\documentclass[12pt]{article}
\DeclareMathAlphabet{\scr}{U}{rsfs}{m}{n}
\usepackage{latexsym,url}
\usepackage[mathscr]{eucal}
\usepackage{amsfonts}
\usepackage{amscd}
\usepackage{cite}         
\usepackage{amssymb}
\usepackage{colordvi}
\usepackage[centertags]{amsmath}
\usepackage{enumerate}
\usepackage{graphicx}
\usepackage{booktabs}

\newcommand{\cleqn}{\setcounter{equation}{0}}
\newcommand{\newc}{\newcommand}
\newc{\be}{\begin{equation}}
\newc{\ee}{\end{equation}}
\newc{\bea}{\begin{eqnarray}}
\newc{\eea}{\end{eqnarray}}
\newc{\ol}{\overline}
\newc{\wt}{\widetilde}
\newc{\bs}{\boldsymbol}
\newc{\m}{\mathcal}

\newc{\mtext}{\Green}

\begin{document}

\title{\hfill ~\\[-30mm]
          \hfill\mbox{\small IPPP-12-40}\\[-3.5mm]
          \hfill\mbox{\small DCPT-12-80}\\[13mm]
       \textbf{
What is the discrete gauge symmetry of the $\bs{R}$-parity violating MSSM?
}}
\date{}
\author{\\
Herbi K. Dreiner\footnote{E-mail: {\tt dreiner@th.physik.uni-bonn.de}}
,~   
Marja Hanussek\footnote{E-mail: {\tt hanussek@th.physik.uni-bonn.de}}
\\[3mm]
  \emph{\small Physikalisches Institut der Universit\"at Bonn,}\\
  \emph{\small Nu{\ss}allee 12, 53115 Bonn, Germany}\\[10mm]
Christoph Luhn\footnote{E-mail: {\tt christoph.luhn@durham.ac.uk}}
\\[3mm]
  \emph{\small Institute for Particle Physics Phenomenology, University of Durham,}\\
  \emph{\small Durham, DH1 3LE, United Kingdom}}

\maketitle

\begin{abstract}
\noindent  
The lack of experimental evidence for supersymmetry motivates
$R$-parity violating realizations of the minimal supersymmetric standard model. 
Dropping $R$-parity, 
alternative symmetries have to be imposed in order to stabilize the
proton. We determine the possible discrete $R$ and non-$R$ symmetries, which
allow for renormalizable $R$-parity violating terms in the superpotential and which, 
at the effective level, are consistent with the constraints from nucleon decay. 
Assuming a gauge origin, we require the symmetry to be discrete gauge
anomaly-free, allowing also for cancellation via the Green Schwarz
mechanism. Furthermore, we demand  lepton-number
  violating neutrino mass terms
either at the renormalizable or nonrenormalizable level. In order to solve
the $\mu$ problem, the discrete $Z_N^{}$ or $Z_N^R$ symmetries have to forbid
any bilinear superpotential operator at tree level.
In the case of renormalizable baryon number violation the smallest possible symmetry
satisfying all conditions is a unique hexality $Z_6^R$. In the case of
renormalizable  lepton-number
  violation the smallest symmetries are two
hexalities, one $Z_6^{}$ and one $Z_6^R$.
\end{abstract}
\thispagestyle{empty}
\vfill
\newpage
\setcounter{page}{1}




\section{Introduction}
\cleqn

The Standard Model (SM) provides a remarkably successful description
of particle physics as observed in past and current experiments. Yet,
it is considered to be incomplete due to the occurrence of quadratic
divergences which directly entail the so-called technical hierarchy
problem~\cite{Gildener:1976ai,Veltman:1980mj}. Various extensions of
the SM have been proposed to cure this problem, the most popular one
being low-scale supersymmetry~\cite{Nilles:1983ge,Martin:1997ns}. However, 
so far no signs of supersymmetry nor, with the exception of massive
neutrinos, {\it any} physics beyond the SM\footnote{Neutrino masses can be elegantly
obtained via the type I seesaw
mechanism~\cite{Minkowski:1977sc,Gell-Mann,Yanagida,Mohapatra} with or without 
supersymmetry by introducing right-handed neutrinos $N^c$.} 
have been seen at the LHC, which severely challenges the minimal supersymmetric
extension of the SM (MSSM)
\cite{Bechtle:2011dm,Allanach:2011qr,Buchmueller:2011sw,Bechtle:2012zk}. 
The lack of experimental evidence for supersymmetry motivates the study of
nonminimal incarnations of supersymmetry, in particular $R$-parity
violating scenarios, see
e.g. \cite{Dreiner:1997uz,Csaki:2011ge,Allanach:2012vj,Dreiner:2012mx,Graham:2012th,Hanussek:2012eh,Dreiner:2012wm}.

Originally, $R$-parity \cite{Farrar:1978xj}, or equivalently matter parity,
was introduced in order to ensure a stable proton
\cite{Dimopoulos:1981dw}. This is achieved by forbidding both baryon~($B$) and
lepton~($L$) number violation at the renormalizable level. Without imposing
$R$-parity, the most general renormalizable superpotential including the SM
particles is given by \cite{Hall:1983id} 
\bea
W&=& y^u Q H_u U^c + y^d Q H_d D^c + y^e L H_d E^c + \mu H_dH_u  \notag\\
&&  + \, \kappa L H_u  + \lambda LLE^c   + \lambda' LQD +  \lambda'' U^cD^cD^c
\ .
\label{renW} 
\eea
Here, $Q$ and $L$ denote the left-handed quark and lepton doublets, while
$U^c$, $D^c$ and $E^c$ correspond to the right-handed fields; $H_u$ and $H_d$
are the up- and the down-type Higgs fields. Note that we have suppressed all
color, weak and family indices. The terms in the second line of
Eq.~\eqref{renW} violate either baryon or lepton number. Forbidding these by
imposing $R$-parity clearly stabilizes the proton in a renormalizable
theory. However, as proton decay requires baryon- as well as  lepton-number
violation, it is equally possible to allow for $L~(B)$ violation provided the $B~(L)$
violating terms are forbidden. This can be achieved by virtue of $R$-parity
violating discrete symmetries such as e.g. baryon
triality~\cite{Ibanez:1991pr,Dreiner:2006xw}.  

In general, alternatives to $R$-parity can be classified according to
\begin{itemize}
\item the allowed renormalizable and nonrenormalizable
  operators\\(i.e. baryon- and  lepton-number  violating dimension three, four,
  five terms; $\mu$~term), 
\item discrete anomaly considerations,\\[0mm] 
  cancellation with or without the Green-Schwarz (GS)
  mechanism~\cite{Green:1984sg},  
\item their compatibility with grand unified theories (GUTs).
\end{itemize}
The idea of constraining possible discrete symmetries using anomaly
considerations was first brought forward by Ib\'a\~nez and
Ross~\cite{Ibanez:1991hv,Ibanez:1991pr}. As global 
discrete symmetries are violated by quantum gravity effects, it is desirable
to obtain them as remnants of a spontaneously broken (continuous) gauge
symmetry~\cite{Krauss:1988zc,Banks:1989ag,Preskill:1990bm,Preskill:1991kd,Banks:1991xj}. In order to be mathematically consistent, the underlying gauge theory  
must be anomaly-free; the corresponding anomaly conditions can then be
translated to (weaker) discrete anomaly conditions which constrain the
low-energy discrete symmetry. Assuming family-independent charges as well as
no exotic light particles, Ib\'a\~nez and Ross studied the anomaly-free $Z_2$
and $Z_3$ symmetries~\cite{Ibanez:1991hv,Ibanez:1991pr}, identifying only two
interesting candidates, matter parity $M_p$ and baryon triality $B_3$. A
subsequent extension of their work to $Z_N$ symmetries with arbitrary values
of $N$ revealed another attractive symmetry, namely proton hexality $P_6$
\cite{Dreiner:2005rd}. As shown in Ref.~\cite{Luhn:2007gq}, adding right-handed
neutrinos to the particle content gives rise to an infinite set of new
discrete symmetries if one assumes neutrinos to be Dirac particles.  All
these discrete symmetries allow for the bilinear term in the first line of
Eq.~\eqref{renW}. As such they do not provide a natural solution to the
$\mu$~problem~\cite{Kim:1983dt} because the dimensionful $\mu$~parameter is
expected to take a value at a scale much higher than the phenomenologically
required electroweak scale.  

A straightforward way to alleviate this state of affairs consists in
forbidding the $\mu$~term with the discrete symmetry. A weak
scale $\mu$ term must then be generated
dynamically~\cite{Giudice:1988yz,Kim:1994eu}.  Adopting the idea of 
eliminating the $\mu$ term through a discrete symmetry, it was
shown in Refs.~\cite{Lee:2010gv,Lee:2011dya} that $SU(5)$ GUT-compatible 
discrete charge assignments are inconsistent with the
discrete anomaly conditions unless discrete symmetries ($Z_N$) are
extended to discrete $R$ symmetries ($Z_N^R$). Requiring $SO(10)$
compatibility, a unique $Z_4^R$ symmetry was identified which forbids
the $\mu$~term as well as dimension three, four and five baryon- and
 lepton-number  violating operators. See also Ref.~\cite{Babu:2002tx}.
When this symmetry is broken to standard matter parity, the $\mu$~term
is generated at the electroweak scale.  For a similar discussion in an
$SU(5)\times U(1)$ setting, see Ref.~\cite{Paraskevas:2012kn}.
Often the GS mechanism is imposed in order for the symmetries to be
consistent with discrete anomaly considerations. For earlier work on $Z_N^R$
symmetries which forbid the $\mu$ term but do not invoke the GS anomaly
cancellation mechanism, see e.g. Refs.~\cite{Kurosawa:2001iq,Hamaguchi:2003za},
where $R$-parity violating operators are suppressed (and absent at the
renormalizable level).  

It is the purpose of this paper to systematically investigate the
$R$-parity violating discrete family-independent $Z_N$ and $Z_N^R$
symmetries which forbid the bilinear terms in Eq.~\eqref{renW}. We
consider discrete symmetries with either $(i)$ renormalizable $B$
violation or $(ii)$ renormalizable $L$ violation, but not both
simultaneously.\footnote{$B$ and $L$ violation in models with discrete
$R$ symmetries which (in the symmetry limit) forbid the $\mu$~term as well as
all renormalizable $B$- and $L$-violating operators was studied e.g. in
Ref.~\cite{Choi:1996fr}.}
These symmetries are further constrained by requiring
the absence of $B$- and/or $L$-violating dimension-five operators
which, if present, would mediate rapid proton decay
\cite{Sakai:1981pk,Weinberg:1981wj,Dimopoulos:1981dw}. In order to
stay as general as possible, we first impose the discrete GS anomaly
cancellation condition only~\cite{Ibanez:1992ji}. The so-obtained
infinite list of discrete symmetries can be significantly reduced by
either demanding anomaly freedom without the GS mechanism or,
alternatively, consistency with the type I seesaw
mechanism~\cite{Minkowski:1977sc,Gell-Mann,Yanagida,Mohapatra}.

The paper is organized as follows. In Sec.~\ref{sec:anomalycond} we present
the discrete anomaly coefficients for $Z_N$ and $Z_N^R$ symmetries and discuss
the resulting anomaly condition invoking the GS mechanism. The
phenomenological constraints on the dimension-five baryon and lepton number
violating operators in the presence of renormalizable $R$-parity violation are
listed in Sec.~\ref{sec:requirements}. Combining these constraints with the GS
anomaly condition in Sec.~\ref{sec:possible}, we obtain all possible
allowed $Z_N^{[R]}$ symmetries.\footnote{We adopt the notation $Z_N^{[R]}$ to
refer to two cases of  discrete $R$ symmetries ($Z_N^R$) and discrete non-$R$
symmetries ($Z_N$); analogously for the continuous symmetries
$U(1)_{[R]}$.} 
This set of viable $R$-parity violating discrete  
symmetries is thinned out by adding further constraints in
Sec.~\ref{sec:further} and Appendix~\ref{app-offensive}.  In
Sec.~\ref{mutermTH} we discuss the implications of dynamically
generating the $\mu$ term. We conclude in Sec.~\ref{sec:conl}. 




\section{\label{sec:anomalycond}Discrete anomaly coefficients}
\cleqn

The discrete anomaly coefficients are derived from the anomaly coefficients of the
underlying gauge theory. We assume this to be the SM gauge group
$SU(3)_C\times SU(2)_W\times U(1)_Y$ augmented by the $U(1)_{[R]}$ gauge
symmetry which gives rise to the discrete $Z_N^{[R]}$
symmetry.\footnote{For gauged $R$ symmetries and their anomalies see 
Refs.~\cite{Freedman:1976uk,Chamseddine:1995gb,Castano:1995ci}.} 
Disregarding the
anomaly coefficients involving only SM factors, we encounter three linear
anomaly coefficients
\be
A_{SU(3)_C-SU(3)_C-U(1)_{[R]}} \ , ~\quad
A_{SU(2)_W-SU(2)_W-U(1)_{[R]}} \ , ~\quad
A_{\mathrm{grav}-\mathrm{grav}-U(1)_{[R]}} \ ,\label{lini}
\ee
where ``grav'' stands for gravity, as well as three purely Abelian anomalies 
\be
A_{U(1)_Y-U(1)_Y-U(1)_{[R]}} \ , ~\quad
A_{U(1)_Y-U(1)_{[R]}-U(1)_{[R]}} \ , ~\quad
A_{U(1)_{[R]}-U(1)_{[R]}-U(1)_{[R]}} \ .\label{abeli}
\ee
We shall not be concerned with the Abelian anomalies in Eq.~\eqref{abeli}
as they are less general~\cite{Preskill:1991kd,Banks:1991xj}. For instance,
the cubic anomaly $A_{U(1)_{[R]}-U(1)_{[R]}-U(1)_{[R]}}$ is derived from the
$U(1)_{[R]}$ charges of all fields, including the massive ones. The
possibility of having fractionally charged heavy particles then allows for solutions to
the cubic anomaly with any $U(1)_{[R]}$ charge assignments for the light
states, see e.g. \cite{Kurosawa:2001iq}. Similarly, the other Abelian
anomalies in Eq.~\eqref{abeli} provide only marginal constraints related to 
heavy fractionally (hyper-)charged particles.\footnote{In models
based on an underlying GUT with a simple Lie algebra, the hypercharge
of all the fields is quantized relative to each other as they
necessarily originate from some GUT multiplet. Such a GUT framework
would render the anomaly coefficient $A_{U(1)_Y-U(1)_Y-U(1)_{[R]}}$
more significant.}  On the other hand, the linear anomalies of
Eq.~\eqref{lini}, lead to severe constraints on the allowed $U(1)_{[R]}$
charge assignments, and thus on the set of possible discrete symmetries. 

In order to formulate the anomaly coefficients for both $U(1)$ and $U(1)_R$
symmetries simultaneously, we follow Ref.~\cite{Lee:2010gv} and introduce the
variable $R$. For $U(1)_R$ symmetries we set $R=1$, while a regular $U(1)$
symmetry has $R=0$. This is necessary since, in supersymmetry, a $U(1)$
symmetry assigns equal charge to the scalar and the spin one-half components
of a chiral superfield, and the components of a vector superfield remain
neutral. The situation is quite different for a $U(1)_R$ symmetry which (depending
on the convention) assigns a charge of $+1$ to the superspace variable
$\theta$. Denoting the $U(1)_R$ charge of a chiral superfield 
\be
\Phi ~=~ \varphi + \theta \psi +\theta^2 F \ ,
\ee
by $x$, the spin one-half component $\psi$, which is the particle
contributing to the anomaly, 
has a charge of $x-1$. Hence, in a unified notation, a chiral superfield with
charge $x$ enters the anomaly coefficients with a factor
of\footnote{Note that a different convention is adopted in 
 Ref.~\cite{Chamseddine:1995gb}.}  
\be
x-R \ , ~\quad \left\{  \begin{array}{l}
R=0 \ , \quad \mathrm{for}~U(1) \ ,\\ 
R=1 \ , \quad \mathrm{for}~U(1)_R \ .\end{array} \right. 
\ee
Concerning $U(1)_{[R]}$ neutral vector superfields, 
\be
V ~=~ \theta \sigma^\mu \bar \theta A_\mu + \theta^2\bar\theta \bar\lambda+
\bar\theta^2 \theta\lambda + \theta^2\bar\theta^2 D \ ,
\ee
which correspond to the gauge fields of the theory, it is clear that the
fermionic components, the gauginos $\lambda$, carry the same $U(1)_{[R]}$
charge as $\theta$, i.e. the charge $R$. 

The linear anomaly coefficients are calculated as the weighted sums of the $U(1)_{[R]}$
charges of all fermions. For the color anomaly we get 
\be
A_{SU(3)_C-SU(3)_C-U(1)_{[R]}} ~=~ \left( \sum_{i=\mathrm{colored}}  \ell ({\bf{r}}_i) \,
(x_i - R) \right) + \ell({\bf 8}) R  \label{a3}
\ ,
\ee
where $\ell ({\bf{r}}_i)$ denotes the Dynkin index of the $SU(3)$
representation ${\bf{r}}_i$ and the sum is over colored chiral states only. These
colored states could be fundamental triplets or higher-dimensional
representations like sextets, octets, etc. The corresponding Dynkin indices are
defined up to an overall normalization. Here, we adopt the standard normalization with
$\ell (\mathrm{fund.})=\frac{1}{2}$. Then the Dynkin index of an octet becomes
$\ell({\bf 8}) = 3$ meaning that the gluinos contribute to the anomaly with
the term $3R$.

In order to derive the {\it discrete} anomaly coefficient from Eq.~\eqref{a3},
we need to relate the $U(1)_{[R]}$ charges $x_i$ to the $Z_N^{[R]}$ charges $q_i$.
Assuming all $x_i$ to be integers\footnote{If they are fractional (but still
quantized), it is possible to rescale the charges by a common factor of~$f$
such that they become integers. However, this will entail the $U(1)_{R}$
charge of $\theta$ to be $f$ rather than 1, potentially leading to more
general sets of discrete $Z_N^R$ symmetries, as pointed out recently in 
Ref.~\cite{MCChen}.}  we can readily express this relation, after
$U(1)_{[R]} \rightarrow Z_N^{[R]}$ breaking, as
\be
x_i ~=~ q_i + m_i N \ ,\label{disc-cont}
\ee
with $m_i \in \mathbb Z$. Inserting this into Eq.~\eqref{a3} we find
\be
A_{SU(3)_C-SU(3)_C-U(1)} ~=~ \left( \sum_{i=\mathrm{colored}}  \ell ({\bf{r}}_i) \,
(q_i - R ) \right) + 3R + \mbox{$\frac{1}{2}$}\,  k \cdot N  \ ,
\label{aa3}
\ee
where the integer $k=\sum_i 2\, \ell({\bf{r}}_i) m_i$ is unspecified in the
low-energy theory. The factor of~$\frac{1}{2}$ arises due to the standard
normalization of the Dynkin indices. At this level, the sum is over light and
heavy fermions alike. Heavy particles, that is particles which decouple from
the low-energy theory and should therefore not occur in any useful discrete
anomaly condition, have a $Z_N^{[R]}$ invariant mass term. This allows us to
remove their contribution from the explicit sum and absorb it into the third
term of Eq.~\eqref{aa3}, proportional to $k$, as we show now. Assuming the
heavy particle to be Dirac entails two independent chiral superfields; their
discrete charges $q_{D_1}$ and $q_{D_2}$ have to add up to $2R~\mathrm{mod}~N$
in order to be compatible with a bilinear mass term in the
superpotential. Therefore, their contribution to the discrete anomaly
coefficient is given by   
\be
\ell ({\bf{r}}_D ) \,(q_{D_1} + q_{D_2} - 2R)  ~=~ \ell ({\bf{r}}_D ) \,k' \,N   \ ,
\ee
where $k' \in\mathbb Z$. 
With $\ell({\bf{r}}_D )$ necessarily being a multiple of $\frac{1}{2}$ it is
clear that such a contribution simply amounts to redefining  the unspecified
parameter $k$ in Eq.~\eqref{aa3}. A heavy Majorana particle, on the other
hand, contributes to the discrete anomaly coefficient with the charge $q_M$ of
only one chiral superfield. The existence of the mass term requires $2q_M =
2R~\mathrm{mod}~N$. Its effect on the anomaly coefficient reads
\be
\ell ({\bf{r}}_M ) \, (q_{M} - R)  ~=~ \ell ({\bf{r}}_M ) \,\mbox{$\frac{1}{2}$} \, k'' \,N  \ ,
\ee
with $k'' \in\mathbb Z$. The Dynkin index of a Majorana particle is
constrained, as the representation must me real. In $SU(3)$ these are the ${\bf
  1,8,27,64,...}$. It is now possible to show that the Dynkin indices of these
and all other real representations are even multiples of $\ell (\mathrm{fund.}) =
\frac{1}{2}$ and thus integers~\cite{Slansky:1981yr}.\footnote{This relies on the
  observation that $\ell({\bf {r}_i} \times {\bf {\ol r}_i}) = d({\bf {r}_i})
  \ell({\bf {\ol r}_i}) + \ell ({\bf {r}_i}) d({\bf {\ol r}_i})  = 2 d({\bf
    {r}_i}) \ell({\bf {r}_i})$, where $d({\bf {r}_i})$ is the dimension of the
  representation.} 
Similar to the case of a heavy Dirac particle, the contribution of a
heavy Majorana particle to the discrete anomaly coefficient 
can therefore be absorbed into the third term of Eq.~\eqref{aa3}.

In summary, the structure of the discrete anomaly coefficient of
Eq.~\eqref{aa3} is unchanged once the heavy particles are removed from
the first term. Then the explicit sum is over the light particles of the
model. Assuming the MSSM particles to be the only light fields 
as well as
family-independent discrete charges, we obtain
\be
A_{SU(3)_C-SU(3)_C-U(1)_{[R]}} ~=~ 3 \cdot  \mbox{$\frac{1}{2}$} \left( 2 q_Q +q_{U^c} + q_{D^c}
-4R  \right) +3R + \mbox{$\frac{1}{2}$}\,  k \cdot N \label{aaa3R}  \ .
\ee
The factor of 3 accounts for the number of families. Similarly one can work
out the discrete anomaly coefficient of the weak anomaly 
\be
A_{SU(2)_W-SU(2)_W-U(1)_{[R]}} ~=~ 3\cdot \mbox{$\frac{1}{2}$} \left( 3 q_Q +q_{L}  -4R  \right)   +
\mbox{$\frac{1}{2}$} \left(  q_{H_u} +q_{H_d} -2R  \right)   +2R  
+ \mbox{$\frac{1}{2}$}\,  \tilde k \cdot N \ ,  \label{aaa2R} 
\ee
where we have assumed one pair of Higgs doublets and $\tilde k\in \mathbb Z$. 

Turning to the gravitational anomaly
$A_{\mathrm{grav}-\mathrm{grav}-U(1)_{[R]}}$, we first remark that it
does not involve any Dynkin indices. We simply need to add the
$U(1)_{[R]}$ charges of all the fermions in the theory, i.e. the
quarks and leptons, the Higgsinos, the gauginos, the gravitino as well
as any additional SM neutral fermions. The MSSM gauginos
(gluino, wino, bino) enter with their multiplicities, $8R+3R+R$, and
the $R$-gaugino, see Ref.~\cite{Chamseddine:1995gb}, adds the term
$R$.\footnote{The latter is not included in Ref.~\cite{Lee:2010gv} as
  the authors envisage a scenario without a local $U(1)_R$ symmetry.}
Furthermore, there is the contribution of the gravitino which adds
$-21R$ to the anomaly coefficient
\cite{Christensen:1978gi,Nielsen:1978ex,Ibanez:1992ji,Castano:1995ci}.
The gravitational discrete anomaly coefficient is then given by 
\bea
A_{\mathrm{grav}-\mathrm{grav}-U(1)_{[R]}} &=& 3 \cdot \left( 6 q_Q
  +3q_{U^c} + 3q_{D^c}+2 q_{L} +q_{E^c} -15R \right)  \notag \\
&& +  \left( 2 q_{H_u} +2 q_{H_d}  -4R \right) -21R+8R+3R+1R+1R ~~~~~~~ \label{aaagravR} \\
&& + \, \hat k \cdot N + \sum_{i=\mathrm{SM~neutral}} q_i \notag \ ,
\eea 
where the integer $\hat k$ originates from the difference between the
$U(1)_{[R]}$ and the $Z_N^{[R]}$ charges, see
Eq.~\eqref{disc-cont}. The sum over SM neutral fermions includes both
heavy and light degrees of freedom. Since the existence of light
hidden fermions is not excluded, this sum can yield an arbitrary
contribution which is not necessarily a half-integer multiple of
$N$. In the following, we will therefore not make use of the
gravitational anomaly to constrain the set of allowed $Z_N^{[R]}$
symmetries.

Invoking the GS mechanism, the discrete anomaly coefficients of
Eqs.~(\ref{aaa3R}) and (\ref{aaa2R}) have to  satisfy the universality
condition 
\be
\frac{A_{SU(3)_C-SU(3)_C-U(1)_{[R]}}}{k_C} =
\frac{A_{SU(2)_W-SU(2)_W-U(1)_{[R]}}}{k_W} = \delta_{\mathrm{GS}} \ ,
\ee
where $\delta_{\mathrm{GS}}\in \mathbb R$  is a constant. Setting this
constant to zero, is tantamount to satisfying the discrete anomaly conditions
without an underlying GS mechanism.
$k_C$ and $k_W$ label the Kac-Moody levels of the corresponding gauge
algebra; they are integers for non-Abelian factors, and furthermore identical
in superstring theories \cite{ramond-kac}. 
In fact, in most string models, the
Kac-Moody levels of the non-Abelian gauge groups are just one. 
We will therefore assume $k_C=k_W$ (see Ref.~\cite{Babu:2002tx} for alternative
choices) so that the relevant discrete anomaly condition reduces to  
\be
3\cdot    \left( q_Q +q_{L} - q_{U^c}-q_{D^c}     \right)   
+ \left(  q_{H_u} +q_{H_d}  \right)   
-4R  
\,= \,
0~\mathrm{mod}~N  \ .\label{gs}
\ee




\section{Phenomenological constraints}\label{sec:requirements}
\cleqn

Besides the anomaly condition in Eq.~\eqref{gs}, the set of allowed $Z_N^{[R]}$ 
symmetries is constrained by various requirements. First,
the discrete symmetry must allow for the up- and the down-type quark as well as
the charged lepton Yukawa terms in Eq.~\eqref{renW}. Second, as pointed out
above, it must forbid the bilinear superpotential terms in order to avoid the $\mu$
(and~$\kappa$) problem. Third, the symmetry should guarantee a sufficiently
stable proton.

Being interested in discrete symmetries which violate
$R$-parity at the renormalizable level, we have to distinguish two (exclusive) cases:
\begin{itemize}
\item[$(i)$] demand renormalizable $B$ violation, i.e. the term $U^cD^cD^c$,
\item[$(ii)$] demand renormalizable $L$ violation, i.e. the term $LLE^c$.
\end{itemize}
Requiring the operator $LLE^c$ in the latter case
entails the existence of the other trilinear $L$-violating term of
Eq.~\eqref{renW}, i.e. $LQD^c$, because the Yukawa operators $LH_dE^c$ and
$QH_dD^c$ are both present. In order to prevent rapid proton decay we need to
forbid renormalizable $L$ violation in case $(i)$, and renormalizable $B$
violation in case $(ii)$. Furthermore, it might be necessary to prohibit some of
the dimension-five $B$- and/or $L$-violating superpotential operators
\cite{Ibanez:1991pr,Allanach:2003eb,Dreiner:2005rd},
\bea
\begin{array}{lllclll}
{\cal O}_1&=&[QQQL]_F\,,& ~~~ & {\cal O}_2 & = & [{U}^c
{U}^c{D}^c{E}^c]_F\,, \\ 
{\cal O}_3 & = & [QQQH_d]_F\,, & ~~~ &   {\cal O}_4 & = & 
[Q{U}^c{E}^cH_d]_F\,, \\ 
{\cal O}_5 & = & [LH_uLH_u]_F\,, & ~~~ & {\cal O}_{6} 
& = & [LH_uH_dH_u]_F\,, \\ 
{\cal O}_{7} & = & [{U}^c{{D}^c}^\ast {E}^c]_D\,, & ~~~ &  
{\cal O}_{8} & = & [{{H}_u}^\ast H_d{E}^c]_D\,, \\ 
{\cal O}_{9} & = & [Q{U}^c{L}^\ast]_D\,, & ~~~ &   {\cal O}_{10} & = & 
[QQ{D^c}^\ast]_D\,,
\end{array} \label{dim5}
\eea
where the subscripts $F$ and $D$ denote the $F$- and $D$-term of the
corresponding product of superfields. Before discussing their role in
destabilizing the proton, it is worthwhile to emphasize that several
of these terms are allowed or forbidden simultaneously. This is due to
the fact that the quark and charged lepton Yukawa terms, the
  first line of Eq.~(\ref{renW}), are necessarily allowed by the
$Z_N^{[R]}$ symmetry. To give an example, let us combine the second
and the complex conjugate of the third term of Eq.~\eqref{renW} and
multiply it by $U^c {U^c}^\ast$. The resulting product is neutral
under $Z_N^{[R]}$. Regrouping it as
\be
QH_dD^c ~ (LH_dE^c)^\ast  ~ U^c{U^c}^\ast ~=~ 
QU^cL^\ast ~ (U^c{D^c}^\ast E^c)^\ast ~ H_d {H_d}^\ast \ ,
\ee
shows that $\mathcal O_7$ and $\mathcal O_9$ have identical $Z_N^{[R]}$ charges.
Similarly it is possible to show that forbidding the bilinear superpotential term $LH_u$
automatically removes the operators $\mathcal O_4$, $\mathcal O_7$, $\mathcal O_8$, 
$\mathcal O_9$. Furthermore, one can easily check that the operators $\mathcal
O_3$ and $\mathcal O_{10}$ are also simultaneously allowed or forbidden.
With these observations, we are well-equipped to discuss the constraints on
the $Z_N^{[R]}$ symmetries originating from the dimension-five operators of
Eq.~\eqref{dim5}. 

The operators $\mathcal O_1$ and $\mathcal O_2$ violate $B$ and $L$ simultaneously; 
hence, they can mediate proton decay without any extra source of $B$ or $L$ violation. 
The operator $QQQL$ needs to be forbidden unless unnaturally small coefficients are 
assumed. On the other hand, the contribution of the operator $U^cU^cD^cE^c$ to 
proton decay is suppressed by small Yukawa couplings~\cite{Sakai:1981pk,Weinberg:1981wj,Dimopoulos:1981dw,Ibanez:1991pr,Murayama:1994tc}.
In the following we will therefore demand the
$Z_N^{[R]}$ symmetry to forbid $\mathcal O_1$ but not necessarily $\mathcal
O_2$. In our list of possible discrete symmetries we will, however, explicitly mark
those cases which allow for the operator $\mathcal O_2$.  

The remaining operators of Eq.~\eqref{dim5} violate either $B$ or $L$, but not
both. In order for them to contribute significantly to proton decay, they need to be
combined with a renormalizable $R$-parity violating operator. Note, however,
that the Weinberg operator $\mathcal O_5$ \cite{Weinberg:1979sa} violates $L$
by two units; it therefore does not yield proton decay even when combined with
the $B$-violating term $U^cD^cD^c$, so it need not be forbidden by the
discrete symmetry. 

The $B$-violating operators $\mathcal O_3$ and $\mathcal O_{10}$ can mediate
proton decay only in case $(ii)$, i.e. the case with renormalizable $L$ violation. The
simultaneous presence of the terms $QH_dD^c$ and $LQD^c$ requires $H_d$ and
$L$ to have identical $Z_N^{[R]}$ charges. As a consequence, forbidding $QQQL$
($\mathcal O_1$) automatically also removes $QQQH_d$  ($\mathcal
O_3$), and with it $\mathcal O_{10}$. Therefore it is not necessary to
separately forbid the operator $\mathcal O_3$.

\begin{figure}
\begin{center}
\includegraphics[height=3cm]{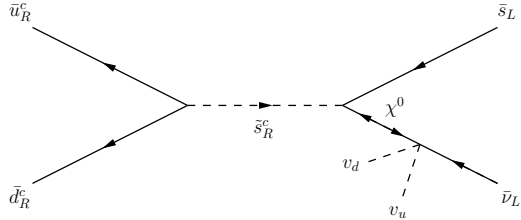}\\[5mm]
\end{center}
\caption{\label{diagram-LHuHdHu}Contribution to proton decay 
obtained from combining $LH_uH_dH_u$~with~$U^cD^cD^c$. Shown here 
is one possible diagram relevant for the decay channel $p \rightarrow  K^+ \bar\nu$.}
\end{figure}

The last dimension-five operator to discuss is $\mathcal O_6$. Violating
lepton number, it has to be combined with the renormalizable term
$U^cD^cD^c$ to mediate proton decay. A possible such diagram is sketched in
Fig.~\ref{diagram-LHuHdHu}. 
Integrating out the neutralino $\chi^0$, the effective couplings on the left-
and right-hand side of the squark propagator multiply to give 
$y_6 \frac{v_u v_d}{M_{\mathrm{Pl}}m_{\chi}} Y^d_{22} \lambda''_{112} $. Here 
we assume the nonrenormalizable operator to be suppressed by the
reduced Planck mass $M_{\mathrm{Pl}}\sim 10^{18}\,\mathrm{GeV}$ and 
denote the corresponding dimensionless coupling constant by $y_6$; in addition, the 
$(2,2)$ element of the down-type quark Yukawa matrix $Y^d$ enters at the
vertex on the right. Comparing this to the better-known bound
$\lambda'_{i22}\lambda''_{112} \lesssim 10^{-27} \left(\frac{\tilde
  m_s}{100\mathrm{GeV}}\right)^2$~\cite{Barbier:2004ez},
obtained from the diagram involving the renormalizable $B$- and $L$-violating
operators of Eq.~\eqref{renW}, yields $y_6 \lambda''_{112} \lesssim 10^{-8}
\left(\frac{\tilde m_s}{100\mathrm{GeV}}\right)^2$ where we have 
used $\frac{v_d Y_{22}^d}{M_{\mathrm{Pl}}} \sim \frac{m_s}{M_{\mathrm{Pl}}}
\sim 10^{-19}$ as well as the assumption $\frac{v_u}{m_{\chi}} \sim 1$.
Even though this bound suggests that one should forbid the operator $\mathcal
O_6$ in case $(i)$, this need not necessarily be the case since the coupling
$\lambda''_{112}$ is already experimentally bounded to be smaller than $10^{-4}$,
but could even be much smaller depending on a hadronic scale parameter
\cite{Goity:1994dq,Allanach:1999ic}.
We therefore do not impose the condition that
$\mathcal O_6$ vanish in case $(i)$, but rather state if a given $Z_N^{[R]}$
symmetry allows for this nonrenormalizable operator or not.

For future reference and convenience, we summarize the constraints on the
$Z_N^{[R]}$ symmetries discussed in this section for both cases,
\begin{itemize}
\item[$(i)$]  with renormalizable $B$ violation:
\begin{itemize}
\item demand existence of $U^cD^cD^c$,
\item forbid $LLE^c$ (thus automatically $LQD^c $),
\item  forbid $H_dH_u$,
\item  forbid $LH_u$ (thus automatically $\mathcal O_{4}$,
  $\mathcal O_{7}$, $\mathcal O_{8}$, $\mathcal O_{9}$),
\item forbid $\mathcal{O}_1=QQQL$;
\end{itemize}
\item[$(ii)$] with renormalizable $L$ violation: 
\begin{itemize}
\item demand existence of $LLE^c$ (thus automatically $LQD^c$),
\item forbid $U^cD^cD^c$,
\item  forbid $H_dH_u$ (thus automatically $LH_u$, $\mathcal O_{4}$,
  $\mathcal O_{7}$, $\mathcal O_{8}$, $\mathcal O_{9}$),
\item forbid $\mathcal{O}_1=QQQL$  (thus automatically $\mathcal O_{3}$ and $\mathcal
  O_{10}$). 
\end{itemize}
\end{itemize}




\section{\label{sec:possible}Possible $\bs{Z_N^{[R]}}$ symmetries}
\cleqn

In this section we combine the phenomenological constraints of the
previous section with the discrete anomaly condition of
Eq.~\eqref{gs}. Including the right-handed neutrino $N^c$,\footnote{Being 
SM gauge singlets, the right-handed neutrinos $N^c$ do not alter the anomaly
coefficients $A_{SU(3)_C-SU(3)_C-U(1)_{[R]}}$ and $A_{SU(2)_W-SU(2)_W-U(1)_{[R]}}$.   
Hence, the results of this section remain valid in scenarios where $N^c$ is
absent.}  we need to fix the generation-independent discrete
$Z_N^{[R]}$ charges of eight superfields. The first set of constraints
on the charge assignments arises from requiring the Yukawa couplings
\be
QH_uU^c \ , \qquad
QH_dD^c \ , \qquad
LH_dE^c \ , \qquad
LH_uN^c \ .\label{sup-allow}
\ee
A further simplification is achieved by shifting all discrete charges by an
amount which is proportional to the hypercharge of the respective
superfield. 
In other words, starting with any given set of discrete charges we can
  obtain a physically equivalent set using such hypercharge shifts.
With the quark doublet having the smallest (absolute value of)
hypercharge, it is possible to set $q_Q=0$ without loss of generality. The
resulting discrete charges take integer values, so that we can 
parameterize the most general $Z_N^{[R]}$ symmetry by
three\footnote{Comparing the number of parameters and constraints we
find: eight discrete charge parameters minus the four constraints of
Eq.~\eqref{sup-allow} minus another degree of freedom related to the
hypercharge shift. This leaves us with three undetermined parameters.} 
integers $p,n,m$ as follows,
\be
\begin{array}{rclcrclcrcl}
q_Q&\!\!=\!\!&0\ ,&~~~~~&q_{U^c}&\!\!=\!\!&-m\ ,&~~~~~&q_{D^c}&\!\!=\!\!&m-n
\ ,\\[2mm]
q_L&\!\!=\!\!&-n-p\ ,&~~~~~&q_{E^c}&\!\!=\!\!&m+p\ ,&~~~~~&q_{N^c}&\!\!=\!\!&-m+n+p
\ ,\\[2mm]
q_{H_u}&\!\!=\!\!&m+2R\ ,&~~~~~&q_{H_d}&\!\!=\!\!&-m+n+2R\ .
\end{array} \label{qs}
\ee
We remark that these discrete charges as well as the parameters $p,n,m$ are only
defined modulo~$N$. For notational simplicity we do not require values between
0 and $N-1$. However, equivalent choices must not be counted separately. 
The parameter $R$ is introduced in the Higgs charges in order to take into
account the possibility of an $R$ symmetry; $R=0$ for a $Z_N$ symmetry, while
$R=1$ in the case of a $Z_N^R$ symmetry. 
Concerning the latter case, we emphasize that the choice of introducing
the parameter $R$ in the Higgs charges, rather than anywhere else, is
completely general, since the other parameters $p,n,m$ have not been fixed at
this stage.

\subsection{Imposing Green-Schwarz anomaly cancellation}

Anomaly considerations further constrain the allowed set of charges in Eq.~\eqref{qs}.
Let us first assume a setup in which the GS anomaly-cancellation mechanism is
at work. Then the anomaly coefficients need not vanish identically. Instead we
only have to satisfy Eq.~\eqref{gs}. Inserting the charges given in
Eq.~\eqref{qs} yields 
\be
n~=~3p ~\mathrm{mod}~N \ ,\label{np}
\ee
independent of $R$. In other words, the parameter $n$ is uniquely determined
by the value of $p$. As we are only interested in symmetries that allow for
renormalizable $B$ or $L$ violation, we can determine the third parameter $m$
from demanding either ($i$) the $B$-violating operator $U^cD^cD^c$  or ($ii$)
the $L$-violating operator $LLE^c$. 
Table~\ref{tab-GSaf} shows the resulting parameterization of $Z_N^{[R]}$
symmetries in the two cases of interest.
\begin{table}
\begin{center}
\begin{tabular}{cc|c|c|c|c}
&&allowed term& $p$ & $n$ & $m$ \\\hline
($i$)\!\!\! &$\not \!\! B$ & $U^cD^cD^c$& $p$ & $3p$ & $6p+2R$\\\hline
($ii$)\!\!\! &$\not \!\! L$ & $LLE^c$& $p$ & $3p$ & $7p+2R$ \\ 
\end{tabular}
\end{center}
\caption{\label{tab-GSaf}The GS anomaly-free $Z^{[R]}_N$ symmetries which
  violate either baryon or lepton number. Here the phenomenological constraints of
  Eqs.~(\ref{restrBV},\ref{restrLV}) have not been taken into account yet.}
\end{table}
Varying $p$ between 0 and $N-1$ shows that, for any given $N$, there are at
most $N$ different $B$-\,/\,$L$-violating $Z_N^{[R]}$ symmetries which allow the
superpotential terms of Eq.~\eqref{sup-allow} and are consistent with the GS
anomaly-cancellation condition of Eq.~\eqref{gs}.

Several of these symmetries additionally allow for superpotential terms which,
if present, would lead to rapid proton decay. As discussed in
Sec.~\ref{sec:requirements},  we need to forbid the operators $LLE^c$,
$H_uH_d$, $H_uL$ and $QQQL$ in the $B$-violating case, leading to the
inequalities  
\be
\begin{array}{ll} (i) ~ \not\!\!B~\text{case}~~~& \left\{
\begin{array}{rcll}
p&\neq & 0~\mathrm{mod}~N                  &~~(LLE^c) 
\ , \\
3p + 2R &\neq & 0~\mathrm{mod}~N      &~~(H_uH_d) 
\ , \\
2p+2R &\neq & 0~\mathrm{mod}~N        &~~(H_uL) 
\ , \\
4p+2R&\neq & 0~\mathrm{mod}~N         &~~(QQQL) 
\ .
\end{array} \right.
\end{array}\label{restrBV}
\ee
Likewise, the operators  $U^cD^cD^c$, $H_uH_d$ and $QQQL$ 
have to be forbidden in the $L$-violating case. Thus, the corresponding
conditions reduce to only three inequalities
\be
\begin{array}{ll} (ii) ~ \not\!\!L~\text{case}~~~& \left\{
\begin{array}{rcll}
p&\neq & 0~\mathrm{mod}~N                    &~~(U^cD^cD^c) 
\ , \\
3p + 2R &\neq & 0~\mathrm{mod}~N        &~~(H_uH_d) 
\ , \\
4p+2R&\neq & 0~\mathrm{mod}~N           &~~(QQQL)
\ .
\end{array} \right.
\end{array}\label{restrLV}
\ee
Notice that the constraints on $p$ in the $L$-violating
case, Eq.~\eqref{restrLV}, must be satisfied in the $B$-violating case as
well. The $B$-violating $Z_N^{[R]}$ symmetries are, however,  additionally
constrained by the third inequality of Eq.~\eqref{restrBV}.


\begin{table}[!t]
\begin{center}
{\footnotesize{
\begin{tabular}{|c|c||c|c|c||c|c|c|} \hline
$N$ & $R$ & $p$ & $n$ & $m$ & anomaly-free & $LH_uLH_u$ & $LH_uH_dH_u$ 
\\  \hline
 5 & 0 & 1 & 3 & 1  &  &  & $\checkmark$ 
  \\
 5 & 0 & 2 & 1 & 2  &  &  & $\checkmark$ 
\\
 5 & 1 & 3 & 4 & 0  &  &  &  
\\ \hline

 6 & 0 & 1 & 3 & 0  &  &  &  
\\
 6 & 1 & 3 & 3 & 2 &  & $\checkmark$ &  
\\ \hline

 7 & 0 & 1 & 3 & 6  &  &  &  
\\
 7 & 0 & 2 & 6 & 5  &  &  &  
\\
 7 & 0 & 3 & 2 & 4  &  &  &  
\\
 
7 & 1 & 1 & 3 & 1 &  &  &  
\\
7 & 1 & 2 & 6 & 0  &  &$\checkmark$  &  
\\
 7 & 1 & 5 & 1 & 4  &  &  &  
\\ \hline

 8 & 0 & 1 & 3 & 6  &  &  &  
\\
 8 & 0 & 3 & 1 & 2 &  &  &  
 \\

 8 & 1 & 1 & 3 & 0  &  &  &  
\\
 8 & 1 & 4 & 4 & 2  &  &  &  
\\
 8 & 1 & 5 & 7 & 0  &  &  &  
 \\
 8 & 1 & 6 & 2 & 6  &  &  &  
\\ \hline

 9 & 0 & 1 & 3 & 6 & $\checkmark$ &  &  
\\
 9 & 0 & 2 & 6 & 3 & $\checkmark$ &  &  
\\
 9 & 0 & 4 & 3 & 6 & $\checkmark$ &  &  
\\

 9 & 1 & 1 & 3 & 8  &  &  &  
\\
 9 & 1 & 2 & 6 & 5  &  &  &  
\\
 9 & 1 & 3 & 0 & 2 &  & $\checkmark$ &  
\\
 9 & 1 & 5 & 6 & 5  &  &  &  
 \\
 9 & 1 & 6 & 0 & 2  &  &  & $\checkmark$ 
 \\
 9 & 1 & 7 & 3 & 8  &  &  &  
 \\ \hline

 10 & 0 & 1 & 3 & 6 &  &  &  
 \\
 10 & 0 & 3 & 9 & 8 &  &  &  
 \\
 
10 & 1 & 1 & 3 & 8 &  & $\checkmark$ &  
 \\
 10 & 1 & 3 & 9 & 0 &  &  & 
 \\
 10 & 1 & 5 & 5 & 2 &  &  &  
  \\
 10 & 1 & 8 & 4 & 0 &  &  &  
\\\hline
 
\end{tabular}
}}
\end{center}
\caption{\label{tab:GS_B}The list of all $B$-violating $Z_N^{[R]}$
  symmetries with $N \leq 10$ which fulfill the Green-Schwarz anomaly
  cancellation condition and the conditions listed in
  Eq.~(\ref{restrBV}). None of these symmetries allows for the superpotential
  term $U^cU^cD^cE^c$.}    
\end{table}

\begin{table}[!p]
\begin{center}
{\footnotesize{
\begin{tabular}{|c|c||c|c|c||c|c|c|} \hline
$N$ & $R$ & $p$ & $n$ & $m$ & anomaly-free & $LH_uLH_u$ & $U^cU^cD^cE^c$ 
\\  \hline
 3 & 1 & 2 & 0 & 1  & $\checkmark$ & $\checkmark$ &   $\checkmark$
 \\ \hline
 
 4 & 1 & 1 & 3 & 1   &  &$\checkmark$  &    $\checkmark$
\\\hline
 
 5 & 0 & 1 & 3 & 2   &  &  &   
\\
 5 & 0 & 2 & 1 & 4   &  &  &   
 \\
 
 5 & 1 & 3 & 4 & 3   &  &  &   
 \\
 5 & 1 & 4 & 2 & 0   &  & $\checkmark$ &    $\checkmark$
\\ \hline

 6 & 0 & 1 & 3 & 1  &  & $\checkmark$ &   
 \\ 

 6 & 1 & 2 & 0 & 4   & $\checkmark$ & $\checkmark$ &   $\checkmark$
 \\
 6 & 1 & 3 & 3 & 5   &  & $\checkmark$ &    
\\ 
 6 & 1 & 5 & 3 & 1   &  & $\checkmark$ &   $\checkmark$
 \\ \hline

 7 & 0 & 1 & 3 & 0   &  &  &   
 \\
 7 & 0 & 2 & 6 & 0   &  &  &   
 \\
 7 & 0 & 3 & 2 & 0   &  &  &    
\\

 7 & 1 & 1 & 3 & 2    &  &  &   
 \\
 7 & 1 & 2 & 6 & 2   &  &  &
 \\
 7 & 1 & 5 & 1 & 2   &  &  &   
 \\
 7 & 1 & 6 & 4 & 2   &  &$\checkmark$  &   $\checkmark$
 \\ \hline

 8 & 0 & 1 & 3 & 7  &  &  &  
  \\ 
 8 & 0 & 3 & 1 & 5   &  &  &   
 \\

 8 & 1 & 1 & 3 & 1   &  &  &  
 \\
 8 & 1 & 3 & 1 & 7   &  &$\checkmark$  &   $\checkmark$
 \\  
 8 & 1 & 4 & 4 & 6   &  &  &  
 \\
 8 & 1 & 5 & 7 & 5  &  &  &   
 \\
 8 & 1 & 6 & 2 & 4   &  &  &    
\\
8 & 1 & 7 & 5 & 3   &  & $\checkmark$ &  $\checkmark$
  \\\hline

 9 & 0 & 1 & 3 & 7   &$\checkmark$  &  &   
 \\
 9 & 0 & 2 & 6 & 5   &$\checkmark$  &  &    
\\
 9 & 0 & 4 & 3 & 1   &$\checkmark$  &  &    
\\

9 & 1 & 1 & 3 & 0   &  &  &
 \\
9 & 1 & 2 & 6 & 7   &  & $\checkmark$ &   
 \\
9 & 1 & 3 & 0 & 5   &  &  &  
  \\
9 & 1 & 5 & 6 & 1   &  & $\checkmark$ &   
 \\
9 & 1 & 6 & 0 & 8  &  & &    
\\
9 & 1 & 7 & 3 & 6   &  &  &    
\\
9 & 1 & 8 & 6 & 4   &  & $\checkmark$ &    $\checkmark$
\\ \hline

 10 & 0 & 1 & 3 & 7   &  &  &   
 \\
 10 & 0 & 3 & 9 & 1   &  &  &   
 \\

10 & 1 & 1 & 3 & 9  &  &  &    
\\ 
10 & 1 & 3 & 9 & 3   &  &  &   
\\
10 & 1 & 4 & 2 & 0   &  & $\checkmark$ & $\checkmark$
   \\
10 & 1 & 5 & 5 & 7   &  &  &   
 \\
10 & 1 & 8 & 4 & 8  &  &  &  
  \\ 
10 & 1 & 9 & 7 & 5   &  & $\checkmark$ &    $\checkmark$
\\\hline
\end{tabular}
}}
\end{center}
\caption{\label{tab:GS_L}The list of all $L$-violating $Z_N^{[R]}$
  symmetries with $N \leq 10$ which fulfill the Green-Schwarz anomaly
  cancellation condition and the conditions listed in Eq.~(\ref{restrLV}).  
As will be discussed in Sec.~\ref{mutermTH}, the symmetries allowing for
$U^cU^cD^cE^c$ are disfavored once the generation of an effective $\mu$ term
is imposed.}   
\end{table}

Taking these restrictions into account it is straightforward to determine the
allowed $Z_N^{[R]}$ symmetries which rely on GS  anomaly-cancellation. We
have listed the smallest ones in Table~\ref{tab:GS_B} for the $B$-violating
and in Table~\ref{tab:GS_L} for the $L$-violating case. For $R=0$
the symmetries defined by $(N,p,n,m)$, $(N,-p,-n,-m)=(N,N-p,N-n,N-m)$ and 
$(\frac{N}{d},\frac{p}{d},\frac{n}{d},\frac{m}{d})$, where $d$ is the greatest
common divisor, are equivalent and therefore only one of them is shown. A
similar overcounting of symmetries does not occur in the case of $R=1$, with,
however, one exception: for $N=4$ the symmetries defined by $(p,n,m)$ and
$(4-p,4-n,4-m)$ are indeed equivalent. This becomes clear by noticing that
$2=\pm 2~\mathrm{mod}~4$. Going from the former symmetry to the latter 
changes {\it all} charges of Eq.~\eqref{qs}, including the charge of the
Higgses. Any superpotential operator which is allowed by one symmetry hence
satisfies the condition 
$$
\sum_i q_i ~=~ 2~\mathrm{mod}~4 ~=~- 2~\mathrm{mod}~4 ~=~ \sum_i (-q_i) \ ,
$$
and, so, is automatically allowed by the other symmetry as well. 
For each of the symmetries listed in Tables~\ref{tab:GS_B} and \ref{tab:GS_L}
we have marked whether or not they are discrete anomaly-free without imposing 
the GS mechanism (see Sec.~\ref{sec-anomalyfree}). Furthermore it is shown
which of the symmetries allow for the Weinberg operator as well as the potentially
critical superpotential operators $LH_uH_dH_u$ (only in the $B$-violating
case) and $U^cU^cD^cE^c$. We point out that the latter carries a $Z_N^{[R]}$
charge of $-2p$ and is thus forbidden by all $B$-violating symmetries that
satisfy the third  inequality of Eq.~\eqref{restrBV}.

It is instructive to see explicitly why there are no $N=2,3,4$
solutions in the $B$-violating case and no $N=2$ solutions in the $L$
violating case.  $N=2$ is excluded by the inequality of
Eqs.~(\ref{restrBV},\ref{restrLV}) related to forbidding the term
$QQQL$ as the left-hand side is always even.  In the $B$-violating
case, non-$R$ symmetries with $N=3$ and $N=4$ are forbidden by the
second and the forth inequality of Eq.~\eqref{restrBV},
respectively. Furthermore, the $B$-violating $Z_3^R$ symmetries with
$p=1$ and $p=2$ are excluded by the fourth and the third inequality of
Eq.~\eqref{restrBV}, respectively. Finally, the $B$-violating $Z_4^R$
symmetries with $p=1$ and $p=2$ are excluded by the third and the
second inequality of Eq.~\eqref{restrBV}, respectively.

\section{\label{sec:further}Further constraints on the list of $\bs{Z_N^{[R]}}$
  symmetries} 
\cleqn
We have seen in the previous section that there exists an infinite number of
both $B$ and $L$-violating symmetries which satisfy the GS anomaly
cancellation condition of Eq.~\eqref{gs}. Several of these are anomaly
free even without invoking the GS mechanism. 
In this section we first identify the complete set of anomaly-free $Z_N^{[R]}$
symmetries parameterized by an integer $\ell$. Another strategy to thin out
the list of possible $Z_N^{[R]}$ symmetries given in
Sec.~\ref{sec:possible} consists of looking for symmetries which are
consistent with the type I seesaw mechanism. Again the obtained solutions can
be parameterized by an integer $\ell$. 

It is worth emphasizing that compatibility with standard grand unified theories cannot
be achieved for $R$-parity violating discrete symmetries which stabilize the
proton. For instance, in $SU(5)$, the renormalizable superpotential operators
$LQD^c$ and $U^cD^cD^c$ both originate in the same $SU(5)$ term, namely ${\bf
  \bar 5\, \bar 5 \, 10}$; allowing for one of the two operators is
accompanied by having the other as well. Therefore, an $SU(5)$
compatible $Z_N^{[R]}$ symmetry would violate both $B$ and $L$ at the
renormalizable level, and the proton would decay
rapidly.\footnote{Another problem which would have to be addressed in GUT
extensions without a $\mu$ term is the existence of massless colored Higgs states.}
Similar arguments hold for $SO(10)$ as well as $SU(5)\times U(1)$. In
contrast, a setup based on Pati-Salam `unification' \cite{Pati:1973uk}
might be consistent with $R$-parity violating $Z_N^{[R]}$ symmetries, however, in
the following we shall not adopt Pati-Salam compatibility as a
constraint on the list of discrete symmetries.

\subsection{\label{sec-anomalyfree}Imposing anomaly freedom}
To further reduce the number of possible discrete symmetries, we can search for
solutions which do not necessarily have to rely on the GS anomaly-cancellation
mechanism. That is, in this subsection we are interested in discrete anomaly
free $Z_N^{[R]}$ symmetries. This adds one more condition to the 
discussion of Sec.~\ref{sec:possible},\footnote{The condition in
Eq.~\eqref{np} ensures that the discrete anomaly coefficients 
$A_{SU(3)_C-SU(3)_C-U(1)_{[R]}} $ and $A_{SU(2)_W-SU(2)_W-U(1)_{[R]}} $ are
identical. Demanding the former (latter) to vanish, automatically also sets
the latter (former) to zero.} to wit
\be
A_{SU(3)_C-SU(3)_C-U(1)_{[R]}} \,=\, 0 \qquad  \Rightarrow \qquad 
3p~=\left( \frac{k_1}{3} \, N -2  R \right) \mathrm{mod}~N \ ,\label{ano-free}
\ee
where we have replaced $n$ by $3p$ using the condition of
Eq.~\eqref{np}. Due to the modulo~$N$ ambiguity it is sufficient to
vary $k_1$ between 0 and 2. With $k_1=0$ we get
$3p+2R=0\,\mathrm{mod}\,N$ which is in contradiction to the
requirement of forbidding the $\mu$ term, see
Eqs.~(\ref{restrBV},\ref{restrLV}). The remaining two cases $k_1=1,2$
yield fractional values for $3p$ unless $N$ is a multiple of~3.
We must therefore assume that 3 divides $N$, 
i.e.  $(3|N)$, which allows us to define an integer $N'$ such that
\be
N~=~3N' \ .
\ee
Using this and expressing the mod~$N$ explicitly by $k_2 N$, we can rewrite
Eq.~\eqref{ano-free} as
\be
p~= \frac{1}{3} (k_1 N' -2 R)  + k_2 N' \ .\label{p-free}
\ee
Since values of $p$ which are identical modulo $N$ are equivalent, we can restrict
ourselves to the cases where $k_2=0,1,2$. Moreover, the fact that $p$ must be
integer requires the factor $(k_1 N' -2 R)$ to be a multiple of 3, which in turn entails
several restrictions and case distinctions:
\begin{itemize}
\item[($a$)] $R=0$ ~$\rightarrow$~ $(3|N')$ ~~$\rightarrow$~~ $N'=3\ell$ \ ,
\item[($b$)] $R=1~\: \land \:~k_1=2$\hspace{0.15mm} ~~$\rightarrow$~~ $N'=1+3\ell$ \ ,
\item[($c$)] $R=1~\: \land \:~k_1=1$\hspace{0.15mm} ~~$\rightarrow$~~ $N'=2+3\ell$ \ ,
\end{itemize}
with $\ell\in \mathbb{N}$.  This shows that the possible discrete symmetries
can be classified according to the value of $N'$. The allowed values of $k_1$
and $k_2$ then give rise to a small set of different $Z_N^{[R]}$
symmetries. For $(b)$ and $(c)$, there are only three different choices,
corresponding to $k_2=0,1,2$. In the case of $(a)$ one can additionally vary
$k_1$, but it is easy to show that the choices with $k_1=2$ are related to
those with $k_1=1$ as we discuss now. Inserting $R=0$ and $N'=3\ell$ into
Eq.~\eqref{p-free} gives
\be
p~=~(k_1   + 3 k_2 )\, \ell  ~\equiv~ c_{k_1,k_2} \, \ell \ ,
\ee
where the coefficient $c_{k_1,k_2}\in\{1,2,4,5,7,8\}$ takes six
  different values (recall $k_1\not=0$). Note, however, that its
  value can always be shifted by a multiple of 9 without affecting the
  obtained discrete symmetry. This is because in case $(a)$
  $N=3N'=9\ell$. Shifting $c_{k_1,k_2}$ by 9 corresponds to shifting
  $p$ by $9\ell$ which is just $N$, and all values are defined mod$\,N$.
For $k_1=1$ we find $c_{k_1,k_2}\in \{1,4,7\}$, while
  for $k_1=2$ we get $\{2,5,8\}=\{-7,-4,-1\}~\mathrm{mod}~9$. The
  overall minus sign between the latter and the former solution is fed
  through from the parameter $p$ to the parameter $n$, Eq.~\eqref{np},
  and eventually -- due to $R=0$ -- also to $m$ and the discrete
  charges $q_i$ of Eq.~\eqref{qs}. Hence, the solutions with $k_1=2$
  are physically identical to the solutions with $k_1=1$, in case $(a)$.

We are now in a position to formulate the most general anomaly-free
$Z_N^{[R]}$ symmetries. We summarize our findings in Table~\ref{tab-classes},
where we have made use of the mod~$N$ ambiguity to simplify the expressions. 
\begin{table}[t]\begin{center}
\begin{tabular}{c|c|c||c|c|c|c}
class&$R$ & $N'$ & $p$ & $n$ & \multicolumn{2}{c}{$m$} \\  \hline
&&&&&$(i)~\not\!\!B$  case &$(ii)~\not\!\!L$ case\\ \hline
$(a)$&0&3&1 & 3  & 6 & 7 \\\cline{4-7}
&&&4 & 3  & 6 & 1 \\\cline{4-7}
&&&7 & 3  & 6 & 4 \\\hline
$(b)$&1&$1+3\ell$&$2\ell$ & $6\ell$  & $-1+3\ell$ & $-1+5\ell$ \\\cline{4-7}
&&&$1+5\ell$ & $6\ell$  & $-1+3\ell$ & $8\ell$ \\\cline{4-7}
&&&$2+8\ell$ & $6\ell$  & $-1+3\ell$ & $-2+2\ell$ \\\hline
$(c)$&1&$2+3\ell$&$\ell$ & $3\ell$  & $2+6\ell$ & $2+7\ell$ \\\cline{4-7}
&&&$2+4\ell$ & $3\ell$  & $2+6\ell$ & $-2+\ell$ \\\cline{4-7}
&&&$4+7\ell$ & $3\ell$  & $2+6\ell$ & $4\ell$ 
\end{tabular}\end{center}
\caption{\label{tab-classes}The list of all anomaly-free discrete $(i)$ baryon and
$(ii)$ lepton number violating $Z_N^{[R]}$ symmetries. Here $N=3N'$
and $\ell\in \mathbb N$ is a free parameter. The parameters $p,n,m$
can be translated to the discrete charges $q_i$ using Eq.~\eqref{qs}.
As discussed in Appendix~\ref{app-offensive}, for $\ell\geq 1$ all
symmetries are consistent with the restrictions of
Eqs.~(\ref{restrBV},\ref{restrLV}).  For $\ell=0$, only two $L$
violating symmetries defined by $p=2$ together with $N'=1,2$ are
phenomenologically viable; however, we show in Sec.~\ref{mutermTH}
that these are disfavored once the generation of an effective $\mu$
term is imposed.}
\end{table}
Notice that for $R=0$, i.e. the case ($a$), the value of $N=9$ is uniquely
specified as a possible overall factor of $\ell$ would only rescale the
charges without changing the physics. For each of the three cases, $(a)$,
$(b)$ and $(c)$, we have three subcases corresponding to the three possible
choices of $k_2=0,1,2$ in Eq.~\eqref{p-free}.

Having found the anomaly-free baryon- and lepton-number violating $Z_N^{[R]}$
symmetries of Table~\ref{tab-classes} it is necessary to identify the subset which
is compatible with the restrictions of Eqs.~(\ref{restrBV},\ref{restrLV}). The
discussion of Appendix~\ref{app-offensive} shows that all symmetries of
Table~\ref{tab-classes} with $\ell \geq 1$ are allowed. With $\ell = 0$, only
two $L$-violating symmetries,  $(R,N,p,n,m)=(1,3,2,0,1)$ and
$(1,6,2,0,4)$, are consistent with
Eqs.~(\ref{restrBV},\ref{restrLV}), while all other symmetries with $\ell=0$
are phenomenologically forbidden. 

The superpotential operators $LH_uH_dH_u$ and $U^cU^cD^cE^c$, which
have the potential to mediate proton decay, can be shown to exist only for a
finite number of anomaly-free $Z_N^R$ symmetries. Imposing the constraints of
Sec.~\ref{sec:possible}, $LH_uH_dH_u$, which has to be considered only in the
$\not\!\!B$ case, carries a $Z_N^{[R]}$ charge of $5p+8R$ [see Eq.~\eqref{qs}
and Table~\ref{tab-GSaf}], so it is only allowed if
\be
5p+6R \,=\,0~\mathrm{mod}~N \ .\label{cond-lhhh}
\ee
Comparing this condition with the results of Table~\ref{tab-classes} shows
that $LH_uH_dH_u$ is only present for the following two $\not\!\!B$ anomaly-free discrete $R$
symmetries $(R,N,p,n,m)=(1,12,6,6,2)$ and $(1,24,18,6,14)$. 
Again using solely the constraints of Sec.~\ref{sec:possible}, the discrete charge of the
second operator, $U^cU^cD^cE^c$, is given by $-2p$,  regardless of $Z_N^{[R]}$
violating $B$ or $L$ at the renormalizable level. Hence, this operator is allowed if
\be
2p+2R \,=\,0~\mathrm{mod}~N \ .\label{cond-uude}
\ee
As this is inconsistent with the third inequality of
Eq.~\eqref{restrBV}, all $B$-violating symmetries forbid
$U^cU^cD^cE^c$. In the $L$-violating case, one can easily show that
the only anomaly-free discrete $R$ symmetries satisfying
Eq.~\eqref{cond-uude} are the two symmetries allowed with $\ell=0$ in
Table~\ref{tab-classes}, i.e. $(R,N,p,n,m)=(1,3,2,0,1)$ and $(1,6,2,0,
4)$.  Anticipating the results of Sec.~\ref{mutermTH}, we point out
that these $L$-violating symmetries are disfavored as the mechanism
which generates an effective $\mu$ term can be adopted to generate an
effective $U^cD^cD^c$ term at a dangerous level.

\subsection{Requiring the seesaw mechanism}
Tables~\ref{tab:GS_B} and \ref{tab:GS_L} show that many of the
phenomenologically viable $Z_N^{[R]}$ symmetries which satisfy the discrete GS
anomaly-cancellation condition forbid the Weinberg operator $LH_uLH_u$. If
allowed, such an operator can naturally be obtained in the framework of the
attractive type I seesaw mechanism; alternatively it can be generated as an effective
nonrenormalizable operator from Planck scale physics. From the symmetry point
of view both options are identical since we have defined the charge of the
right-handed neutrinos $N^c$ such that the Dirac Yukawa term $LH_uN^c$ is
allowed. As a consequence, demanding the existence of the right-handed Majorana
mass term $N^cN^c$, is equivalent to demanding the Weinberg operator
$LH_uLH_u$. 
In this subsection we will constrain the set of $Z_N^{[R]}$ symmetries found in
Sec.~\ref{sec:possible} by requiring the presence of the Weinberg
operator. Thus  we identify those symmetries which are consistent with the
type I seesaw mechanism. We first discuss the $B$-violating case and turn to
the $L$-violating case thereafter.

The absence of $LH_uLH_u$ in the $B$-violating case entails that neutrinos are
Dirac particles, a scenario which would be ruled out if neutrinoless double
beta decay was observed. This motivates the extraction of only those $Z_N^{[R]}$
symmetries of Table~\ref{tab:GS_B} and its extension to arbitrary $N$ which
are consistent with Majorana neutrinos, i.e. which allow for $LH_uLH_u$. 
All possible $B$-violating $Z_N^{[R]}$ symmetries which satisfy Eq.~\eqref{gs}
are parameterized in terms of three integers $(R,N,p)$,
cf. Table~\ref{tab-GSaf}. Demanding the Weinberg operator yields the condition
\be
2 \,(m-n-p+2R) -2R 
~=~
4p+6R 
~=~ 0 ~\mathrm{mod}~N \ .
\ee
It is clear that $R=0$ conflicts with the absence of the term $QQQL$, see
Eq.~\eqref{restrBV}. We are therefore left with $R=1$, leading to the condition
\be
2p ~=\,\left(  
-3 +\frac{k_3}{2} N   
\right) \,\mathrm{mod} ~N \ , \label{eq:pconstraintMaj}
\ee
with $k_3=0,1$. As $p$ is defined to be an integer, the two cases allow only
particular values of~$N$,
\begin{itemize}
\item[($a$)] ~$
k_3=0$\hspace{0.15mm} ~~$\rightarrow$~~ $N=3+2\ell$ \ ,
\item[($b$)] ~$
k_3=1$\hspace{0.15mm} ~~$\rightarrow$~~ $N=2\,(1+2\ell)$ \ ,
\end{itemize}
where $\ell \in \mathbb N$. Inserting this into Eq.~\eqref{eq:pconstraintMaj}
determines $p$ as a function of $\ell$. In case $(a)$, we find $p=\ell$,
leading to the set of symmetries defined by 
\be
(a) \quad \rightarrow \quad (R,N,p,n,m) ~=~ (1\,,\,3+2\ell\,,\, \ell\,,\, -3+\ell \,,\, -4+2\ell) \ .
\ee
It is straightforward to show that consistency with the restrictions of
Eq.~\eqref{restrBV} requires $\ell \geq 2$. The choice $\ell=0$ would allow
$LLE^c$, while $\ell=1$ would allow $H_uH_d$. Turning to case $(b)$ we obtain
two possible solutions, $p=-1+\ell$ as well as $p=3\ell$, leading to 
\be
(b) \quad \rightarrow \quad (R,N,p,n,m) ~=
\left\{\begin{array}{l}
 (1\,,\,2+4\ell\,,\, -1 + \ell\,,\, -3+3\ell \,,\, -6+2\ell) \ , \\[3mm]
 (1\,,\,2+4\ell\,,\, 3 \ell\,,\, -4+\ell \,,\, -6+2\ell) \ .
\end{array}\right.
\ee
Again it is possible to verify that all such symmetries with $\ell \geq 3$ are
consistent with the restrictions of Eq.~\eqref{restrBV}; moreover, $\ell=2$ is
allowed in the first subcase (with $p=-1+\ell$) and $\ell=1$ is possible in
the second (with $p=3 \ell$). The choice $\ell=0$ would allow $H_uL$ in both
subcases, while $\ell=1$ would allow $LLE^c$ in the first subcase, and choosing
$\ell=2$ in the second would yield the $\mu$ term $H_uH_d$.
We summarize our findings of the allowed GS anomaly-free $B$-violating $Z_N^{[R]}$
symmetries which are consistent with the type I seesaw mechanism, and thus light
Majorana neutrinos, in Table~\ref{tab:BVmaj}. One can quickly confirm that
these solutions do not satisfy Eq.~\eqref{cond-lhhh}, so that $LH_uH_dH_u$ is
forbidden. $U^cU^cD^cE^c$ is forbidden even without demanding the Weinberg
operator due to the incompatibility of Eq.~\eqref{cond-uude} with the third
inequality of Eq.~\eqref{restrBV}. 
\begin{table}[t]\begin{center}
\begin{tabular}{c|c|c||c|c|c||c}
class&$R$ & $N$ & $p$ & $n$ & $m$ & allowed values of  $\ell$  \\  \hline
$(a)$&1&$3+2\ell$ &$\ell$ & $-3+\ell$  & $-4+2\ell$ & $ \ell \geq 2$ \\\hline
$(b)$&1&$2+4\ell$&$-1+\ell$ & $-3+3\ell$  & $-6+2\ell$ & $\ell\geq 2$ \\\cline{4-7}
&&&$3\ell$ & $-4+\ell$  & $-6+2\ell$ & $\ell=1 \,,\, \ell \geq 3$ 
\end{tabular}\end{center}
\caption{\label{tab:BVmaj}The list of all GS anomaly-free $B$-violating $Z_N^{[R]}$
symmetries consistent with the type I seesaw mechanism as well as the
restrictions of Eq.~\eqref{restrBV}.
Here $\ell\in \mathbb N$, and the discrete  charges $q_i$ of the MSSM
superfields are obtained from the parameters $p,n,m$ using Eq.~\eqref{qs}.}  
\end{table}

In the case of $L$-violating discrete $Z^{[R]}_N$ symmetries, the
left-handed neutrinos can acquire a mass radiatively without the
assumption of an underlying seesaw mechanism
\cite{Hall:1983id,Grossman:1997is,Allanach:2011de,Dreiner:2011ft}. Yet,
it is conceivable that such loop-induced contributions to the neutrino
masses are too small to account for the observed lower mass bound of
the heaviest neutrino, and that a seesaw mechanism might still be
required. In that context, it is interesting to extract those GS
anomaly-free $L$-violating $Z_N^{[R]}$ symmetries which allow for the
Weinberg operator. We therefore proceed analogously to the $B$
violating case. The existence of $LH_uLH_u$ gives rise to the
constraint \be 2 \,(m-n-p+2R) -2R ~=~ 6p+6R ~=~ 0 ~\mathrm{mod}~N \
.\label{eq:L-maj} \ee With $0<p+R\leq N$, there are six possible cases
to distinguish, \be p ~=~ \frac{k_4}{6} \, N - R \ ,\label{eq:L-maj2}
\ee with $k_4=1,...,6$. For $k_4=1,5$, it is obvious that $N$ has to
be a multiple of $6$, so we can define an integer $\ell \in \mathbb N$
such that $N=6\ell$; inserting this into Eq.~\eqref{eq:L-maj2} yields
\be p ~=~ k'_4 \,\ell - R \ ,\label{eq:L-maj3} \ee where
$k'_4=k_4=1,5$. Similarly, for $k_4=2,4$, $N$ must be divisible by
$3$, so we can define an integer $\ell$ such that $N=3\ell$, leading
to Eq.~\eqref{eq:L-maj3}, but now with $k'_4=\frac{k_4}{2}=1,2$.
Likewise, $k_4=3$ requires $N=2\ell$, giving Eq.~\eqref{eq:L-maj3}
with $k'_4=\frac{k_4}{3}$.  Finally, $k_4=6$ corresponds to $N=\ell$
and results in Eq.~\eqref{eq:L-maj3} with
$k'_4=\frac{k_4}{6}=1$. These solutions have to be compared to the
phenomenological constraints of Eq.~\eqref{restrLV}. For $R=0$ we
quickly find that $k_4$ can be either $1$ or $5$. After rescaling of
charges and dropping overall signs, we only find one $L$-violating
discrete symmetry with $R=0$. In the cases where $R=1$, one can show
that all symmetries are consistent with Eq.~\eqref{restrLV} provided
that $\ell\geq 3$; some symmetries satisfy the restrictions of
Eq.~\eqref{restrLV} also for $\ell=2$. Our results for the GS anomaly
free $L$-violating $Z_N^{[R]}$ which allow for the Weinberg operator
are summarized in Table~\ref{tab:LVmaj}, with the last column giving
the constraints on $\ell$ arising form the phenomenological
restrictions of Eq.~\eqref{restrLV}.  It is straightforward to prove
that all $Z_N^R$ symmetries obtained from the second ($N=\ell$) and
the third ($N=2\ell$) row of Table~\ref{tab:LVmaj} satisfy
Eq.~\eqref{cond-uude}, thus allowing for the superpotential operator
$U^cU^cD^cE^c$, while the remaining solutions of Table~\ref{tab:LVmaj}
eliminate this term. Note that $L$-violating symmetries which allow
for $U^cU^cD^cE^c$ are disfavored, see Sec.~\ref{mutermTH}.

\begin{table}[t]\begin{center}
\begin{tabular}{c|c||c|c|c||c}
$R$ & $N$ & $p$ & $n$ & $m$  & allowed values of $\ell$  \\  \hline
0&$6$ &$1$ & $3$  & $1$  \\\hline
1&$\ell$ &$-1+\ell$ & $-3+\ell$ & $-5+\ell$ & $ \ell \geq 3$  \\\hline
1&$2\ell$ &$-1+\ell$ & $-3+\ell$ & $-5+\ell$ & $ \ell \geq 2$  \\\hline
1&$3\ell$ &$-1+\ell$ & $-3+3\ell$ & $-5+\ell$ & $ \ell \geq 3$  \\\cline{3-6}
1& &$-1+2\ell$ & $-3+3\ell$ & $-5+2\ell$ & $ \ell \geq 2$  \\\hline
1&$6\ell$ &$-1+\ell$ & $-3+3\ell$ & $-5+\ell$ & $ \ell \geq 2$  \\\cline{3-6}
1& &$-1+5\ell$ & $-3+3\ell$ & $-5+5\ell$ & $ \ell \geq 2$  
\end{tabular}\end{center}
\caption{\label{tab:LVmaj}The list of all GS anomaly-free $L$-violating $Z_N^{[R]}$
symmetries consistent with the type I seesaw mechanism as well as the
restrictions of Eq.~\eqref{restrLV}.  Here $\ell\in \mathbb N$, and
the discrete charges $q_i$ of the MSSM superfields are obtained from
the parameters $p,n,m$ using Eq.~\eqref{qs}.  The $Z_N^R$ symmetries
with $N=\ell$ and $N=2\ell$ allow for $U^cU^cD^cE^c$ and are thus
disfavored once the generation of an effective $\mu$ term is imposed,
see Sec.~\ref{mutermTH}.}
\end{table}

Before concluding this section, we comment on the possibility of combining the 
compatibility of the $Z_N^{[R]}$ symmetries with the type I seesaw mechanism
and the requirement of anomaly freedom, i.e. the results of Sec.~\ref{sec-anomalyfree}.
It is straightforward to verify that there are only two such discrete
symmetries in each case,
\begin{itemize}
\item[$(i)$]  ~$\not \!\!B$ case: ~ $(R,N,p,n,m)= (1,15,6,3,8)
  ~\mathrm{and}~(1,30,6,18,8)$  ,
\item[$(ii)$] ~$\not \!\!L$ case: ~ $(R,N,p,n,m)= (1,3,2,0,1)
  ~\mathrm{and}~(1,6,2,0,4)$  ,
\end{itemize}
where the $L$-violating $Z_N^R$ symmetries might lead to rapid proton
decay as the same mechanism which generates an effective $\mu$
term can be adopted to generate an effective $U^cD^cD^c$ term, see
Sec.~\ref{mutermTH}.




\section{Consequences of generating an effective $\bs{\mu}$ term}
\label{mutermTH}
\cleqn

From the low-energy perspective, a $\mu$ term at around the electroweak scale
is mandatory. Having forbidden this term by the $Z_N^{[R]}$ symmetry, we need to
generate it dynamically by breaking $Z_N^{[R]}$ spontaneously. In this
section we will discuss the ensuing phenomenological consequences.

Regardless of the explicit mechanism which is responsible for creating an effective
$\mu$ term in the superpotential, it must necessarily break $Z_N^{[R]}$. Let
us for concreteness assume the Giudice-Masiero
mechanism~\cite{Giudice:1988yz} in which the bilinear term $H_dH_u$ is
obtained from the nonrenormalizable K\"ahler potential operator
$\frac{S^\dagger}{M_{\mathrm{Pl}}} H_dH_u$. When the $F$-term of $S$ acquires
a vacuum expectation value  $\langle F_S \rangle \sim m_{3/2}
M_{\mathrm{Pl}}$, with $m_{3/2}$ denoting the gravitino mass, the term  
\be
m_{3/2} H_dH_u \ ,\label{muterm}
\ee
is generated in the effective superpotential after integrating out the
superspace variable~${\bar \theta}^2$. 
Using Eq.~\eqref{qs}, the $Z_N^{[R]}$ charge of this effective term is
$n+4R\neq 2R\,\mathrm{mod}\,N$. Before $Z_N^{[R]}$ breaking, the field which
gives rise to this effective $\mu$ parameter carries a discrete charge such that
$m_{3/2}$ can be regarded as an object with charge $-n-2R$.

In principle, trilinear terms can be generated by the same mechanism, the only
difference being an extra $\frac{1}{M_{\mathrm{Pl}}}$ suppression of the
operator. In the $B$-violating case $(i)$, the crucial operator to look at is
$LQD^c$; if this was generated analogously to the $\mu$ term, we would get
\be
\frac{m_{3/2}}{M_{\mathrm{Pl}}} LQD^c \ .\label{lqdterm}
\ee
Assuming $m_{3/2} \sim 100\,\mathrm{GeV}$, this corresponds numerically to 
$\lambda' \sim 10^{-16}$. This, together with the presence of $U^cD^cD^c$ at
the renormalizable level, could lead to proton decay at a dangerous
rate. Eq.~\eqref{lqdterm} does, however, not occur if the charge of $LQD^c$
and the effective charge of $m_{3/2}$ (i.e. $-n-2R$) add up to something
different from $2R\,\mathrm{mod}\,N$. Explicitly, we find using Eq.~\eqref{qs}
and Table~\ref{tab-GSaf},  
\be
m-2n-p-n-2R =-4p ~\neq ~2R~\mathrm{mod}~N \ .
\ee
As this condition is identical to the fourth inequality of Eq.~\eqref{restrBV}, the $L$
violating renormalizable term $LQD^c$ cannot be obtained in the $B$-violating
case $(i)$ by the same mechanism that gives rise to the effective $\mu$ term. 

Similar considerations lead to the condition 
\be
-2p ~\neq ~2R~\mathrm{mod}~N \ ,\label{mu+uud}
\ee
on the $L$-violating $Z_N^{[R]}$ symmetries which allow for an effective $\mu$
(and thus also $\kappa$) term and, at the same time, do not generate the $B$
violating term $U^cD^cD^c$.  Notice that the condition of Eq.~\eqref{mu+uud} is
equivalent to forbidding the superpotential term $U^cU^cD^cE^c$,
cf.~Eq.~\eqref{cond-uude}. Hence all $L$-violating symmetries which allow for
$U^cU^cD^cE^c$, see Table~\ref{tab:GS_L} as well as comments below 
Eqs.~\eqref{cond-uude} and \eqref{eq:L-maj3}, are disfavored by imposing the
generation of an effective $\mu$ term. 

We conclude this section by pointing out that the original $Z_N^{[R]}$ symmetry
is, in many cases, not completely broken through the mechanism which generates the
effective $\mu$ term. There exists a simple criterion for having a residual
symmetry: $N$ and $n+2R$ (i.e. the absolute value of the effective charge of
$m_{3/2}$) must have a common divisor. Denoting the greatest common divisor
by $M$, a $Z_N^{[R]}$ symmetry is broken to a $Z_M^{[R]}$ symmetry. In the case
of the anomaly-free discrete symmetries listed in Table~\ref{tab-classes} one can
easily show that $M=N'$. As another example we mention the $L$-violating $Z_6$
(non-$R$) symmetry of Table~\ref{tab:GS_L}, defined by $(p,n,m)=(1,3,1)$. With
$n+2R=3$ we get $M=3$, so that the residual $Z_3$ symmetry is given by 
$(p',n',m')=(p,n,m)\,\mathrm{mod}\,3 = (1,0,1)$, which is the well-known 
symmetry baryon triality~$B_3$. 




\section{\label{sec:conl}Discussion and Conclusion}
\cleqn

The parameter space of the conventional $R$-parity conserving MSSM is becoming
ever more constrained by the ongoing searches for supersymmetry at the LHC.
The fact that no signal has yet been found sets quite stringent
bounds on the masses of some strongly interacting sparticles.
In particular, first generation squarks and gluinos below about 1.5~TeV are
excluded if their masses are roughly equal. On the other hand, squark and
gluino masses above 1.5~TeV seem already somewhat high, considering that the
main motivation for postulating their existence is to stabilize the
electroweak hierarchy against radiative corrections.
However, these mass limits can be evaded in alternative supersymmetric models
such as the $R$-parity violating MSSM, where the lightest supersymmetric
particle decays and thus the missing transverse momentum is considerably
reduced compared to the $R$-parity conserving MSSM. 

In the framework of the $R$-parity violating MSSM, we have identified the (GS
and non-GS) anomaly-free discrete gauge ($R$ and non-$R$) symmetries which are
consistent with constraints from nucleon decay and which, at the same time,
forbid the $\mu$ term. An effective $\mu$ term of electroweak order must then
be generated dynamically via a mechanism such as the one proposed by  Giudice
and Masiero or Kim and Nilles. Furthermore, we consider which symmetries allow
for neutrino mass generation via the Weinberg operator $LH_uLH_u$, or
equivalently which allow for the type I seesaw mechanism if right-handed
neutrinos are added to the particle spectrum. 

As the simultaneous presence of renormalizable $B$-violating and $L$-violating
terms is disfavored because it would lead to rapid proton decay, we have analyzed
the two cases separately.
In the case of renormalizable $B$ violation we find exactly two anomaly-free
discrete gauge symmetries which allow for the Weinberg operator:
a $Z_{15}^{R}$ and a $Z_{30}^{R}$, given at the end of Sec.~\ref{sec:further}. 
Relaxing the constraints by imposing anomaly-cancellation via the
Green-Schwarz mechanism, we also find solutions with smaller values of $N$, the
smallest being a unique $B$-violating hexality $Z_6^{R}$ defined by
$(p,n,m)=(3,3,2)$, cf.~Eq.~(\ref{qs}).\footnote{There is an
infinite set of solutions with larger $N$.} The required dynamical
generation of the $\mu$ term entails the breaking of this~$Z_6^{R}$, leaving
no residual symmetry at all.  

In the $L$-violating case, there are exactly two anomaly-free discrete gauge
symmetries which allow for the Weinberg operator: one with $N=3$ and one with
$N=6$, given at the end of Sec.~\ref{sec:further}. However these are
disfavored due to constraints from proton decay as the mechanism which
generates the effective $\mu$ term can be adopted to generate an effective
$U^cD^cD^c$ term at a dangerous level (which generally happens for symmetries
which do not forbid the term $U^cU^cD^cE^c$). Therefore we again extend the set of
symmetries by imposing anomaly-cancellation via the Green-Schwarz
mechanism. Then, the smallest solutions are the two $L$-violating hexalities $Z_6$ with
$(p,n,m)=(1,3,1)$ and $Z_6^{R}$ with $(p,n,m)=(3,3,5)$. The generation of the
effective $\mu$ term breaks the $Z_6$ down to baryon triality ($B_3$), while
the $Z^R_6$ is broken down to nothing.
Alternatively, one can give up the presence of the Weinberg operator, since,
in the $L$-violating case, sufficiently large neutrino masses can also be
generated radiatively via the dimension-four $LQD$ and $LLE$ operators.  In
that case, the smallest viable anomaly-free solutions are the three $L$-violating
ennealities\footnote{For Greek prefixes see for example
\url{http://en.wikipedia.org/wiki/Number_prefix.}} 
$Z_9$ of Table~\ref{tab:GS_L}, which all reduce to $B_3$ once the $\mu$ term 
is generated.




\section*{Acknowledgments}

H.D. would like to thank Graham Ross for discussions and the
SFB TR-33 The Dark Universe for financial support.
M.H. would like to thank the Deutsche Telekom Stiftung and the
Bonn-Cologne Graduate School of Physics for financial support.
C.L. acknowledges support from the  EU ITN grants UNILHC 237920 and 
INVISIBLES 289442, and thanks the Physics Institute in Bonn for hospitality.




\section*{Appendix}

\begin{appendix}

\section{\label{app-offensive}Phenomenologically viable anomaly-free
  $\bs{Z^{[R]}_N}$} 

In Sec.~\ref{sec-anomalyfree} we have derived all anomaly-free $B$ or $L$
violating ${Z^{[R]}_N}$ symmetries regardless of their phenomenological
viability. It is the purpose of this appendix to compare the obtained
solutions with the constraints of Eqs.~(\ref{restrBV},\ref{restrLV}), and 
identify those symmetries which are physically relevant. To do so, we tabulate
the discrete charges of the offensive operators in Table~\ref{tab-offensive}
for each of the nine cases defined by the choice of $(R,N',p)$,
cf. Table~\ref{tab-classes}. Clearly, in each case the charges only depend on
the integer parameter $\ell$. Using the mod~$N$ ambiguity we have shifted all
charges such that the coefficient of $\ell$ is always positive and smaller than
nine. This way it is straightforward to verify that, for $\ell\geq 1$,  the
charges of the offensive operators are always non-zero and smaller than
$N$. Hence, all symmetries with $\ell\neq 0$ are consistent with
Eqs.~(\ref{restrBV},\ref{restrLV}).  For $\ell=0$, there exist several entries
in Table~\ref{tab-offensive} which vanish modulo $N$. Hence, with $\ell=0$
(i.e. $N=3,6$) there is no solution which forbids {\it all} dangerous
operators of the $B$-violating case $(i)$. However, in the $L$-violating case
$(ii)$ we do not have to consider the last column of Table~\ref{tab-offensive}; as a
consequence, the $L$-violating symmetries with $(R,N,p,n,m)=(1,3,2,0,1)$ and
$(1,6,2,0,4)$ are possible.

\begin{table}[t]\begin{center}
\begin{tabular}{c|c|c||c|c|c|c}
\multicolumn{3}{r||}{$(i)$ operators in $\not\!\!B$  case} &$LLE^c$
&$H_uH_d$ &$QQQL$& $H_uL$ \\ \hline
\multicolumn{3}{r||}{$(ii)$ operators in $\not\!\!L$  case}
&$U^cD^cD^c$&$H_uH_d$&$QQQL$& $-$\\ \hline
$R$ & $N'$ & $p$ & $p$ & $3p+2R$  & $4p+2R$ & $2p+2R$ \\  \hline\hline
0&3&1 & 1  & 3 & 4 & 2 \\\cline{3-7}
&&4 & 4  & 3 & 7 & 8 \\\cline{3-7}
&&7 & 7  & 3 & 1 & 5  \\\hline
1&$1+3\ell$&$2\ell$ & $2\ell$ & $2+ 6\ell$  & $2+8\ell$ &$2+4\ell$\\\cline{3-7}
&&$1+5\ell$ & $1+5\ell$ & $2+6\ell$  & $2\ell$ & $1+\ell$ \\\cline{3-7}
&&$2+8\ell$ & $2+8\ell$ & $2+6\ell$  & $1+5\ell$  &$3+7\ell$ \\\hline
1&$2+3\ell$&$\ell$ & $\ell$ & $2+3\ell$  & $2+4\ell$ & $2+2\ell$ \\\cline{3-7}
&&$2+4\ell$ & $2+4\ell$ & $2+3\ell$  & $4+7\ell$ & $6+8\ell$\\\cline{3-7}
&&$4+7\ell$ & $4+7\ell$ & $2+3\ell$  & $\ell$ & $4+5\ell$
\end{tabular}\end{center}
\caption{\label{tab-offensive}The discrete charges of the offensive operators
  in terms of the parameter $\ell\in \mathbb N$. Here we have used the mod~$N$
  ambiguity to shift the charges such that the coefficient of $\ell$ is
  positive and smaller than nine. The last column need not  be considered in
  the $L$-violating case $(ii)$.} 
\end{table}

\end{appendix}





\begin{thebibliography}{99}

\bibitem{Gildener:1976ai}
  E.~Gildener,
  Phys.\ Rev.\ D {\bf 14} (1976) 1667.

\bibitem{Veltman:1980mj}
  M.~J.~G.~Veltman,
  Acta Phys.\ Polon.\ B {\bf 12} (1981) 437.

\bibitem{Nilles:1983ge}
  H.~P.~Nilles,
  Phys.\ Rept.\  {\bf 110} (1984) 1.

\bibitem{Martin:1997ns}
  S.~P.~Martin,
  hep-ph/9709356.

\bibitem{Minkowski:1977sc}
  P.~Minkowski,
  Phys.\ Lett.\ B {\bf 67} (1977) 421.

\bibitem{Gell-Mann}
M.~Gell-Mann, P.~Ramond, and R.~Slansky,
in Sanibel Talk, CALT-68-709, Feb 1979; hep-ph/9809459 (retroprint);
and in {\em Supergravity} (North-Holland, Amsterdam 1979).

\bibitem{Yanagida}
T.~Yanagida,
in {\em Proceedings of the Workshop on Unified Theory and Baryon
  Number of the Universe}, KEK, Japan, Feb 1979.

\bibitem{Mohapatra}
R.~N.~Mohapatra and G.~Senjanovic,
  Phys.\ Rev.\ Lett.\ {\bf 44} (1980) 912.

\bibitem{Bechtle:2011dm}
  P.~Bechtle, B.~Sarrazin, K.~Desch, H.~K.~Dreiner, P.~Wienemann, M.~Kramer, C.~Robens and B.~O'Leary,
  Phys.\ Rev.\ D {\bf 84} (2011) 011701
  [arXiv:1102.4693].

\bibitem{Allanach:2011qr}
  B.~C.~Allanach, T.~J.~Khoo and K.~Sakurai,
  JHEP {\bf 1111} (2011) 132
  [arXiv:1110.1119].

\bibitem{Buchmueller:2011sw}
  O.~Buchm\"uller, R.~Cavanaugh, A.~De Roeck, M.~J.~Dolan, J.~R.~Ellis, H.~Flacher, S.~Heinemeyer and G.~Isidori {\it et al.},
  Eur.\ Phys.\ J.\ C {\bf 72} (2012) 1878
  [arXiv:1110.3568].
  
\bibitem{Bechtle:2012zk}
  P.~Bechtle, T.~Bringmann, K.~Desch, H.~Dreiner, M.~Hamer, C.~Hensel, M.~Kramer and N.~Nguyen {\it et al.},
  JHEP {\bf 1206} (2012) 098
  [arXiv:1204.4199 [hep-ph]].

\bibitem{Dreiner:1997uz}
  H.~K.~Dreiner,
  in Kane, G.L. (ed.): Perspectives on supersymmetry II 565
  [hep-ph/9707435].

\bibitem{Csaki:2011ge}
  C.~Csaki, Y.~Grossman and B.~Heidenreich,
  Phys.\ Rev.\ D {\bf 85} (2012) 095009
  [arXiv:1111.1239].

\bibitem{Allanach:2012vj}
  B.~C.~Allanach and B.~Gripaios,
  JHEP {\bf 1205} (2012) 062
  [arXiv:1202.6616].

\bibitem{Dreiner:2012mx}
  H.~K.~Dreiner, K.~Nickel, F.~Staub and A.~Vicente,
  Phys.\ Rev.\ D {\bf 86} (2012) 015003
  [arXiv:1204.5925 [hep-ph]].

\bibitem{Graham:2012th}
  P.~W.~Graham, D.~E.~Kaplan, S.~Rajendran and P.~Saraswat,
  JHEP {\bf 1207} (2012) 149
  [arXiv:1204.6038 [hep-ph]].

\bibitem{Hanussek:2012eh}
  M.~Hanussek and J.~S.~Kim,
  Phys.\ Rev.\ D {\bf 85} (2012) 115021
  [arXiv:1205.0019 [hep-ph]].

\bibitem{Dreiner:2012wm}
  H.~Dreiner, W.~Porod, F.~Staub and A.~Vicente,
  arXiv:1205.0557.

\bibitem{Farrar:1978xj}
  G.~R.~Farrar and P.~Fayet,
  Phys.\ Lett.\ B {\bf 76} (1978) 575.

\bibitem{Dimopoulos:1981dw}
  S.~Dimopoulos, S.~Raby and F.~Wilczek,
  Phys.\ Lett.\ B {\bf 112} (1982) 133.

\bibitem{Hall:1983id}
  L.~J.~Hall and M.~Suzuki,
  Nucl.\ Phys.\ B {\bf 231} (1984) 419.

\bibitem{Ibanez:1991pr}
  L.~E.~Ib\'a\~nez and G.~G.~Ross,
  Nucl.\ Phys.\ B {\bf 368} (1992) 3.

\bibitem{Dreiner:2006xw}
  H.~K.~Dreiner, C.~Luhn, H.~Murayama and M.~Thormeier,
  Nucl.\ Phys.\ B {\bf 774} (2007) 127
  [hep-ph/0610026].

\bibitem{Green:1984sg}
  M.~B.~Green and J.~H.~Schwarz,
  Phys.\ Lett.\ B {\bf 149} (1984) 117.

\bibitem{Ibanez:1991hv}
  L.~E.~Ib\'a\~nez and G.~G.~Ross,
  Phys.\ Lett.\ B {\bf 260} (1991) 291.

\bibitem{Krauss:1988zc}
  L.~M.~Krauss and F.~Wilczek,
  Phys.\ Rev.\ Lett.\  {\bf 62} (1989) 1221.

\bibitem{Banks:1989ag}
  T.~Banks,
  Nucl.\ Phys.\ B {\bf 323} (1989) 90.

\bibitem{Preskill:1990bm}
  J.~Preskill and L.~M.~Krauss,
  Nucl.\ Phys.\ B {\bf 341} (1990) 50.

\bibitem{Preskill:1991kd}
  J.~Preskill, S.~P.~Trivedi, F.~Wilczek and M.~B.~Wise,
  Nucl.\ Phys.\ B {\bf 363} (1991) 207.

\bibitem{Banks:1991xj}
  T.~Banks and M.~Dine,
  Phys.\ Rev.\ D {\bf 45} (1992) 1424
  [hep-th/9109045].

\bibitem{Dreiner:2005rd}
  H.~K.~Dreiner, C.~Luhn and M.~Thormeier,
  Phys.\ Rev.\  D {\bf 73} (2006) 075007
  [hep-ph/0512163].

\bibitem{Luhn:2007gq}
  C.~Luhn and M.~Thormeier,
  Phys.\ Rev.\  D {\bf 77} (2008) 056002
  [arXiv:0711.0756].

\bibitem{Kim:1983dt}
  J.~E.~Kim and H.~P.~Nilles,
  Phys.\ Lett.\ B {\bf 138} (1984) 150.

\bibitem{Giudice:1988yz}
  G.~F.~Giudice and A.~Masiero,
  Phys.\ Lett.\ B {\bf 206} (1988) 480.

\bibitem{Kim:1994eu}
  J.~E.~Kim and H.~P.~Nilles,
  Mod.\ Phys.\ Lett.\ A {\bf 9} (1994) 3575
  [hep-ph/9406296].

\bibitem{Lee:2010gv}
  H.~M.~Lee, S.~Raby, M.~Ratz, G.~G.~Ross, R.~Schieren, K.~Schmidt-Hoberg and P.~K.~S.~Vaudrevange,
  Phys.\ Lett.\  B {\bf 694} (2011) 491
  [arXiv:1009.0905].

\bibitem{Lee:2011dya}
  H.~M.~Lee, S.~Raby, M.~Ratz, G.~G.~Ross, R.~Schieren, K.~Schmidt-Hoberg and P.~K.~S.~Vaudrevange,
  Nucl.\ Phys.\  B {\bf 850} (2011) 1
  [arXiv:1102.3595].

\bibitem{Babu:2002tx}
  K.~S.~Babu, I.~Gogoladze and K.~Wang,
  Nucl.\ Phys.\  B {\bf 660} (2003) 322
  [hep-ph/0212245].

\bibitem{Paraskevas:2012kn}
  M.~Paraskevas and K.~Tamvakis,
  Phys.\ Rev.\ D {\bf 86} (2012) 015009
  [arXiv:1205.1391 [hep-ph]].

\bibitem{Kurosawa:2001iq}
  K.~Kurosawa, N.~Maru and T.~Yanagida,
  Phys.\ Lett.\ B {\bf 512} (2001) 203
  [hep-ph/0105136].

\bibitem{Hamaguchi:2003za}
  K.~Hamaguchi and N.~Maru,
  Phys.\ Rev.\ D {\bf 67} (2003) 115003
  [hep-ph/0302163].

\bibitem{Choi:1996fr}
  K.~Choi, E.~J.~Chun and H.~D.~Kim,
  Phys.\ Rev.\ D {\bf 55} (1997) 7010
  [hep-ph/9610504].

\bibitem{Sakai:1981pk}
  N.~Sakai and T.~Yanagida,
  Nucl.\ Phys.\ B {\bf 197} (1982) 533.

\bibitem{Weinberg:1981wj}
  S.~Weinberg,
  Phys.\ Rev.\ D {\bf 26} (1982) 287.

\bibitem{Ibanez:1992ji}
  L.~E.~Ib\'a\~nez,
  Nucl.\ Phys.\  B {\bf 398} (1993) 301
  [hep-ph/9210211].

\bibitem{Freedman:1976uk}
  D.~Z.~Freedman,
  Phys.\ Rev.\ D {\bf 15} (1977) 1173.

\bibitem{Chamseddine:1995gb}
  A.~H.~Chamseddine and H.~K.~Dreiner,
  Nucl.\ Phys.\  B {\bf 458} (1996) 65
  [hep-ph/9504337].

\bibitem{Castano:1995ci}
  D.~J.~Casta\~no, D.~Z.~Freedman and C.~Manuel,
  Nucl.\ Phys.\ B {\bf 461} (1996) 50
  [hep-ph/9507397].

\bibitem{MCChen}
  M.-C.~Chen, M.~Ratz, C.~Staudt, P.~K.~S.~Vaudrevange, 
  arXiv:1206.5375.

\bibitem{Slansky:1981yr}
  R.~Slansky,
  Phys.\ Rept.\  {\bf 79} (1981) 1.

\bibitem{Christensen:1978gi}
  S.~M.~Christensen and M.~J.~Duff,
  Phys.\ Lett.\ B {\bf 76} (1978) 571.

\bibitem{Nielsen:1978ex}
  N.~K.~Nielsen, M.~T.~Grisaru, H.~R\"omer and P.~van Nieuwenhuizen,
  Nucl.\ Phys.\ B {\bf 140} (1978) 477.

\bibitem{ramond-kac}
  P.~Ramond,
  in proceedings of the SUSY~95 conference, Ecole Polytechnique, Palaiseau, France; 
  preprint UFIFT-HEP-95-27.
  
\bibitem{Allanach:2003eb}
  B.~C.~Allanach, A.~Dedes and H.~K.~Dreiner,
  Phys.\ Rev.\ D {\bf 69} (2004) 115002
   [Erratum-ibid.\ D {\bf 72} (2005) 079902]
  [hep-ph/0309196].

\bibitem{Murayama:1994tc}
  H.~Murayama, D.~B.~Kaplan,
  Phys.\ Lett.\  {\bf B336 } (1994)  221
  [hep-ph/9406423].

\bibitem{Weinberg:1979sa}
  S.~Weinberg,
  Phys.\ Rev.\ Lett.\  {\bf 43} (1979) 1566.

\bibitem{Barbier:2004ez}
  R.~Barbier, C.~Berat, M.~Besancon, M.~Chemtob, A.~Deandrea, E.~Dudas, P.~Fayet and S.~Lavignac {\it et al.},
  Phys.\ Rept.\  {\bf 420} (2005) 1
  [hep-ph/0406039].

\bibitem{Goity:1994dq}
  J.~L.~Goity and M.~Sher,
  Phys.\ Lett.\ B {\bf 346} (1995) 69, Erratum-ibid.\ B {\bf 385} (1996) 500
  [hep-ph/9412208].

\bibitem{Allanach:1999ic}
  B.~C.~Allanach, A.~Dedes and H.~K.~Dreiner,
  Phys.\ Rev.\ D {\bf 60} (1999) 075014
  [hep-ph/9906209].

\bibitem{Pati:1973uk}
  J.~C.~Pati and A.~Salam,
  Phys.\ Rev.\ D {\bf 8} (1973) 1240.

\bibitem{Grossman:1997is}
  Y.~Grossman and H.~E.~Haber,
  Phys.\ Rev.\ Lett.\  {\bf 78} (1997) 3438
  [hep-ph/9702421].

\bibitem{Dreiner:2011ft}
  H.~K.~Dreiner, M.~Hanussek, J.~-S.~Kim and C.~H.~Kom,
  Phys.\ Rev.\ D {\bf 84} (2011) 113005
  [arXiv:1106.4338].

\bibitem{Allanach:2011de}
  B.~C.~Allanach, C.~H.~Kom and M.~Hanussek,
  Comput.\ Phys.\ Commun.\  {\bf 183} (2012) 785
  [arXiv:1109.3735].


\end{thebibliography}
\end{document}